\documentclass{jpp}
\usepackage{graphicx}
\usepackage{subcaption}
\usepackage[hidelinks]{hyperref}
\usepackage{academicons}
\usepackage{xcolor}
\usepackage{amsmath}
\usepackage{bm}

\hypersetup{
  colorlinks   = true, 
  urlcolor     = blue, 
  linkcolor    = blue, 
  citecolor   = blue 
}

\usepackage{scalerel,tikz}
\usetikzlibrary{svg.path}
\definecolor{orcidlogocol}{HTML}{A6CE39}
\tikzset{orcidlogo/.pic={\fill[orcidlogocol] svg{M256,128c0,70.7-57.3,128-128,128C57.3,256,0,198.7,0,128C0,57.3,57.3,0,128,0C198.7,0,256,57.3,256,128z}; \fill[white] svg{M86.3,186.2H70.9V79.1h15.4v48.4V186.2z} svg{M108.9,79.1h41.6c39.6,0,57,28.3,57,53.6c0,27.5-21.5,53.6-56.8,53.6h-41.8V79.1z M124.3,172.4h24.5c34.9,0,42.9-26.5,42.9-39.7c0-21.5-13.7-39.7-43.7-39.7h-23.7V172.4z} svg{M88.7,56.8c0,5.5-4.5,10.1-10.1,10.1c-5.6,0-10.1-4.6-10.1-10.1c0-5.6,4.5-10.1,10.1-10.1C84.2,46.7,88.7,51.3,88.7,56.8z};}}
\newcommand\orcidicon[1]{\href{https://orcid.org/#1}{\mbox{\scalerel*{
\begin{tikzpicture}[yscale=-1,transform shape]\pic{orcidlogo};
\end{tikzpicture}}{|}}}}

\usepackage[utf8]{inputenc}
\usepackage[T1]{fontenc}

\newcommand{\theintegral}{David~N.~Hosking}

\newcommand{\xb}{\bm{x}}
\newcommand{\rb}{\bm{r}}
\newcommand{\vb}{\bm{u}}
\newcommand{\bb}{\bm{B}}
\newcommand{\ab}{\bm{A}}
\newcommand{\Ab}{\bm{A}}
\newcommand{\Bb}{\bm{B}}
\newcommand{\Sb}{\bm{s}}

\newcommand{\Fb}{\bm{F}}

\newcommand{\cs}{c_\mathrm{s}}
\newcommand{\va}{v_\mathrm{A0}}

\newcommand{\emag}{\mathcal{E}_\mathrm{M}}
\newcommand{\lmag}{\xi_\mathrm{M}}
\newcommand{\lmagi}{\xi_\mathrm{M0}}

\newcommand{\prandtl}{\mathrm{Pm}}
\newcommand{\lundquist}{S}
\newcommand{\mach}{\mathrm{Ma}}
\newcommand{\blue}[1]{\textcolor{black}{#1}}

\shorttitle{Conservation of magnetic-helicity fluctuations due to decorrelation of fluxes}
\shortauthor{J.~K.~J. Hew, D.~N. Hosking, C. Federrath,  J.~R. Beattie \& N. Kriel}

\title{Conservation of magnetic-helicity fluctuations due to spatial decorrelation of fluxes in decaying MHD turbulence}

\author{Justin Kin Jun Hew\aff{\orcidicon{0000-0002-5238-6115}\,1,2}, 
  \theintegral\aff{\orcidicon{0000-0002-7958-6993}\,3,4,5}\corresp{\email{dnh26@cam.ac.uk}},  Christoph~Federrath\aff{\orcidicon{0000-0002-0706-2306}\,1}, James R. Beattie\aff{\orcidicon{0000-0001-9199-7771}\,6,7} \& Neco~Kriel\aff{\orcidicon{0000-0002-3558-3926}\,1} }

\affiliation{\aff{1}Research School of Astronomy and Astrophysics, Australian National University, Canberra, ACT 2611, Australia
\aff{2}Australia’s Climate Simulator (ACCESS-NRI), Australian National University, Canberra, ACT 2601, Australia
\aff{3}Princeton Center for Theoretical Science, Princeton, NJ 08540, USA
\aff{4}Department of Applied Mathematics and Theoretical Physics, University of Cambridge, Centre for Mathematical Sciences, Wilberforce Road, Cambridge, CB3 0WA, UK
\aff{5}Gonville \& Caius College, Trinity Street, Cambridge, CB2 1TA, UK
\aff{6}Department of Astrophysical Sciences, 112 Nassau Street, Princeton University, Princeton, NJ 08544, USA
\aff{7}Canadian Institute for Theoretical Astrophysics, University of Toronto, 60 St. George Street, Toronto, ON M5S 3H8, Canada
}

\begin{document}

\maketitle

\begin{abstract}
\cite{HoskingSchekochihin20decay} have proposed that \blue{statistically isotropic} decaying MHD turbulence without net magnetic helicity conserves the mean square fluctuation level of magnetic helicity in large volumes---or, equivalently, the integral over space of the two-point correlation function of the magnetic-helicity density, denoted~$I_H$. Formally, the conservation and gauge invariance of $I_H$ require the vanishing of certain boundary terms related to the strength of long-range spatial correlations. These boundary terms represent the ability (or otherwise) of the turbulence to organise fluxes over arbitrarily large distances to deplete or enhance fluctuations of magnetic helicity. In this work, we present a theory of these boundary terms, \blue{employing a methodology analogous to that of}
\citet{BatchelorProudman56} \blue{to determine the relevant asymptotic forms of correlation functions}. We find that \blue{long-range correlations of sufficient strength} to violate the conservation of~$I_H$ \blue{cannot} develop dynamically \blue{if the evolution equation for the magnetic vector potential is chosen to be local in space. Likewise, we find that such correlations cannot develop for a wide class of gauge choices that make this equation non-local (including the Coulomb gauge)}. Nonetheless, we also identify a class of non-local gauge \blue{choices} for which correlations that are sufficiently strong to violate the conservation of $I_H$ do appear possible. We verify our theoretical predictions for the case of the Coulomb gauge with measurements of correlation functions in a high-resolution numerical simulation.
\end{abstract}

\section{Introduction \label{sec:introduction}}
Decaying turbulence is ubiquitous in astrophysical \citep[e.g.,][]{mac1998kinetic,porter1994kolmogorov, BanerjeeJedamzik04}, geophysical \citep[e.g.,][]{metais1986statistical} and engineering \citep[e.g.,][]{kang2003decaying} flows. Theories of decaying turbulence often employ the idea of \emph{selective decay}, i.e., that a quantity better conserved than energy---a \emph{rugged invariant}---requires scaling relations that constrain the dependence of energy on time~\citep{Kolmogorov41c, MatthaeusMontgomery80}. Canonical examples of such conserved quantities in hydrodynamics include the Saffman \citep{saffman67} and Loitsyanksy \citep{Loitsyansky39} integrals---see~\cite{Davidson15} for a review.

In magnetohydrodynamics (MHD), theories of selective decay may be constructed using rugged invariants associated with the magnetic field. In particular, the magnetic helicity (or topological linking number,~\citealt{Moffatt69}),
\begin{equation}
    H = \int_{\blue{\mathbb{R}^3}} d^3 \xb \, \ab\bcdot\bb,\label{helicity}
\end{equation}
of the magnetic field $\bb=\mathbf{\nabla}\times \ab$ is conserved as energy decays if the magnetic Reynolds number is large~\citep{Berger84}. \citet{Woltjer58a} showed that linear force-free magnetic fields (i.e., those for which $\mathbf{\nabla}\times\bb=\alpha \bb$ with $\alpha$ constant) extremise magnetic energy subject to the fixed total magnetic helicity \eqref{helicity}, a fact utilised by \citet{taylor1974} to explain the insensitivity to initial conditions of relaxed magnetic fields in fusion experiments (see~\citealt{TaylorNewton15} for a review). Later, \cite{Hatori84} used the conservation of magnetic helicity to predict \blue{self-similar-decay laws for MHD turbulence arising from an initially helical and statistically isotropic magnetic-field configuration}: Combining ${B^2 \xi_M\sim \mathrm{const}}$ from~\eqref{helicity} with the (seemingly) dimensionally inevitable ${\xi_M\sim B t}$---where $B$ is the characteristic magnetic-field strength in units of the Alfv\'en speed, $\xi_M$ the integral or ``energy-containing'' scale and $t$ time---leads to the predictions ${B^2\propto t^{-2/3}}$ and ${\xi_M\propto t^{2/3}}$. These scalings have been confirmed with direct numerical simulations~\citep{BiskampMuller99,BiskampMuller00, Brandenburg17}, modulo certain caveats about the role of slow magnetic reconnection in changing the decay rate~\citep{Zhou19,Bhat21, HoskingSchekochihin20decay}. 

The selective-decay phenomenology was extended to non-helical magnetic fields, i.e., those for which $H$ vanishes globally, by \citet{HoskingSchekochihin20decay}. They argued that such turbulence conserves an integral similar in form to Saffman's \citep{saffman67} but involving magnetic helicity, viz.,
\begin{equation}
    I_H = \int_{\blue{\mathbb{R}^3}} d^3 \rb \, \langle h(\xb) h (\xb + \rb) \rangle, \label{I_H}
\end{equation}
where $h\equiv \ab \bcdot \bb$ is the magnetic-helicity density, $\xb$ a spatial coordinate, $\rb$ a spatial separation and angle brackets denote an ensemble average.\footnote{\citet{HoskingSchekochihin20decay} refer to $I_H$ as the ``Saffman helicity invariant''. \citet{Zhou22_Hosking} and subsequent studies have referred to $I_H$ as the ``Hosking integral'' for symmetry with the names of the Saffman and Loitsyanksy integrals, and to avoid the suggestion that $I_H$ is parity dependent (see the discussion in \citealt{Brandenburg23b}).} \blue{Intuition for the physical meaning of $I_H$ may be obtained by} replacing the ensemble average in~\eqref{I_H} by a spatial average over a large control volume \blue{$V\gg \xi_M^3$ (i.e., assuming that spatially distant magnetic structures can be considered different realisations of the ensemble)}, which yields
\begin{equation}
    I_H = \lim_{V\to \infty} \frac{1}{V}\left[\int_V d^3 \xb \, h(\xb)\right]^2 .\label{I_Hfluctuation}
\end{equation}
Equation~\eqref{I_Hfluctuation} indicates that $I_H$ is a measure of the random fluctuation level of magnetic helicity that arises in volumes much larger than the integral scale $\xi_M$. 

Conservation of $I_H$ follows formally from the continuity equation for magnetic helicity:
\begin{equation}
    \frac{\partial h}{\partial t} + \bnabla\bcdot \Fb = \blue{D_{\eta}}, \quad \Fb =  \vb(\ab \bcdot \bb) - \bb(\ab \bcdot \vb) - \bb\zeta+\blue{\Fb_{\eta}},\label{helicityflux}
\end{equation}where $\Fb$ is the magnetic-helicity flux, $\vb$ the fluid velocity, $D_{\eta}\equiv -2\eta\bb\bcdot (\bnabla \times \bb)$ the resistive dissipation of magnetic helicity, $\Fb_{\eta} \equiv- \eta \ab\times (\bnabla \times \bb)$ the resistive flux of magnetic helicity, $\eta$ the magnetic diffusivity
and $\zeta$ the gauge function, defined by
\begin{equation}
    {\frac{\partial \ab}{\partial t} =\vb\times \bb}+\bnabla\zeta - \eta \bnabla \times \bb.\label{dAdt}
\end{equation}
\blue{Using~\eqref{helicityflux} to form an equation for $d I_{H}/dt$, and employing the simplifications of statistical isotropy and reflectional symmetry, we obtain}
\begin{equation}
    \frac{d I_H}{d t} = C_{\infty} + \blue{2\int_{\mathbb{R}^3} d^3 \rb \langle D_{\eta}(\xb)h(\xb + \rb)\rangle},\label{dIHdt}
\end{equation}where 
\begin{align}
    C_{\infty} & \equiv  8\pi\lim_{r\to\infty} r^2 \langle \Fb(\xb)h(\xb+\rb)\rangle \bcdot\hat{\rb} \nonumber\\ & = 8\pi \lim_{r\to\infty}r^2\left(\langle u_iA_jB_jA_l'B_l'\rangle  -\langle u_jA_jB_iA_l'B_l'\rangle  -  \langle \zeta B_i A_l'B_l'\rangle + \blue{\langle (\Fb_{\eta})_ {i}h'\rangle}\right)\hat{r}_i,\label{Cinfty}
\end{align}where $r=|\bm{r}|$ and $\hat{\rb}=\rb/r$ is a unit vector in the $\rb$ direction. In the second line, we adopt a shorthand in which unprimed fields are evaluated at $\xb$ and primed ones at $\xb + \rb$. \blue{The resistive terms in~\eqref{dIHdt} and~\eqref{Cinfty} vanish as $\eta \to 0$ (i.e., as the Lundquist number $\lundquist=v_A \xi_M/\eta\to\infty$).\footnotemark~
$I_H$ is then conserved if $C_{\infty}=0$.} 

\citet{HoskingSchekochihin20decay} assumed that~$C_{\infty}$ vanishes. In this case, $I_H \sim B^4 \xi_M^5$ is conserved, which, combined with $\xi_M\sim Bt$, implies the non-helical decay laws 
\begin{equation}
    B^2 \propto t^{-10/9},\quad \xi_M\propto t^{4/9}.\label{nonhelicallaws}
\end{equation}The laws~\eqref{nonhelicallaws} agree with the numerical simulations presented by \cite{HoskingSchekochihin20decay}, who also measured $I_H$ directly and found it to be conserved. \cite{Zhou22_Hosking} verified these results with independent simulations, while corollaries and extensions of the theory have also been confirmed numerically by~\cite{Brandenburg23c}, \cite{Brandenburg23a},  \cite{Brandenburg25_columnar} and \cite{Brandenburg25_twoconserved}.

\footnotetext{\blue{We justify this with the following simple scaling argument. First, we note that the rate of magnetic-energy dissipation via ohmic diffusion, $\mathcal{R}_{\eta}\equiv -\langle \eta |\bnabla \times \bb|^2 \rangle$, cannot diverge as $\eta \to 0$. Dimensionally, $\mathcal{R}_{\eta}\sim f_{\eta} \eta \langle |\bb|^2 \rangle /\ell_{\eta}^2$, where $f_\eta$ is the volume-filling fraction of regions in which dissipation acts---this decreases with decreasing $\eta$ if dissipation is intermittent---and $\ell_\eta$ is the typical diffusive scale (i.e., $|\bnabla \times \bb|\sim \langle |\bb|^2 \rangle^{1/2}/\ell_{\eta}$), which also decreases with decreasing $\eta$. As $\eta \to 0$, the integral on the right-hand side of~\eqref{dIHdt} is $\sim f_{\eta}\xi_M^3 (\eta \langle |\bb|^2 \rangle/\ell_\eta)\langle h^2\rangle^{1/2} \sim \langle |\bb|^2 \rangle\xi_M^4 \ell_{\eta} \mathcal{R}_{\eta}\to 0$. Likewise, $|\langle \Fb_{\eta}h' \rangle| \lesssim \langle |\Fb_{\eta}|\rangle \langle|h| \rangle \sim f_{\eta}\langle |\bb|^2 \rangle^2 \xi_M^2 \eta /\ell_{\eta}\sim \langle |\bb|^2 \rangle \xi_M^2\ell_{\eta}\mathcal{R}_{\eta}\to 0$ in~\eqref{Cinfty}. We conclude that there is no resistive evolution of $I_H$ on $\eta$-independent timescales as $\eta\to 0$.}}

The goal of this paper is to address from first principles whether $C_{\infty}$ vanishes. For it not to do so, helicity fluctuations would need to become sufficiently correlated with helicity fluxes to enhance or diminish the mean square fluctuation level of magnetic helicity [right-hand side of~\eqref{I_Hfluctuation}]. This could be the case if, for example, helicity fluctuations tended to attract or repel each other non-locally. There are two non-local effects that could enable interactions between distant points. First, in subsonic turbulence, distant fluid elements communicate via the pressure field to \blue{reduce compressive motions}. In absolutely incompressible flow (zero Mach number), this is a non-local effect. Whether pressure-mediated interactions violate the conservation of the Loitsyansky integral in incompressible hydrodynamic turbulence, via a boundary term analogous to $C_{\infty}$, remains an unsolved problem~\citep{davidson2000loitsyansky, Ishida06}. 

Secondly,~\eqref{dAdt} is non-local for choices of gauge for which \blue{$\zeta$ is non-local, i.e., depends on integrals of $\vb$, $\bb$ and/or $\ab$ over space (these include the Coulomb gauge $\bnabla \bcdot \ab = 0$)}.  For such gauges,~\eqref{dAdt} can correlate $\boldsymbol{A}$ at distant points.  Plausibly,  therefore,  $I_H$ is conserved only when evaluated in gauges that are, in some sense, sufficiently local (see Section~\ref{sec:conclusion} for discussion of how such a situation might be interpreted physically). \blue{Evidently, $I_H$ would not be a gauge-invariant quantity if its conservation depended on the choice of gauge, and, indeed,  
it may be shown (see Appendix~\ref{app:gauge-IH})} that $I_H$ is gauge invariant only if long-range correlations are sufficiently weak: Under $\ab\to \ab + \mathbf{\nabla}\varphi$,
\begin{align}
    I_H & \to  I_H + \int_{\blue{\mathbb{R}^3}} d^3\rb \, \left[\frac{\partial}{\partial r_j}\langle A_i B_i \varphi^\prime B_j^\prime\rangle -\frac{\partial}{\partial r_i}\langle \varphi  B_i A_j^\prime B_j^\prime\rangle- \frac{\partial^2}{\partial r_i \partial r_j}\langle \varphi  B_i \varphi^\prime B_j^\prime\rangle\right]\nonumber\\
    &= I_H - 4\pi\lim_{r\to\infty} \left[ 2\blue{r^2}\theta + r^2 \frac{d f}{d r} + \frac{d}{dr}\left(r^2 g\right)\right],\label{I_Hgaugetransformation}
\end{align}
where 
\begin{equation}
    \theta (r) = \langle \varphi B_i A_j^\prime B_j' \rangle\hat{r}_i,\quad \langle \varphi  B_i \varphi^\prime B_j^\prime\rangle = f(r)\delta_{ij} + g(r)\hat{r}_i \hat{r}_j.
\end{equation}$I_H$ is unchanged by the gauge transformation only if the limit on the right-hand side vanishes, as it does if distant points are statistically independent.\footnote{\citet{Zhou22_Hosking} calculated $I_H$ in simulations of decaying turbulence for two different choices of gauge (Coulomb and resistive) and found it to be the same for both (see their Figure~2).} 

A summary of our paper and its main results is as follows. In Section~\ref{sec:invariance_perservation}, we \blue{review the sources of long-range correlations in isotropic MHD turbulence. In Section~\ref{sec:calculation}}, we determine the large-$r$ asymptotic behaviour of the correlation functions that appear in~\eqref{Cinfty} by examining their Taylor series in time evaluated at the initial instant, and identifying the dominant terms at large $r$ (as~\citealt{BatchelorProudman56} did for hydrodynamic turbulence). We find that $C_{\infty}$ vanishes for any gauge for which $\zeta$ is a local function \blue{of $\vb$, $\bb$ and $\ab$}, or for which $\zeta$ satisfies a Poisson equation $\nabla^2 \zeta = \phi$, with $\phi$ a local function of $\vb$ and $\bb$. This broad class of gauges contains all those that are commonly used to analyse MHD turbulence, including the Coulomb gauge, for which \blue{${\phi = -\bnabla \bcdot (\vb \times \bb)}$}. Interestingly, we do not find that $I_H$ is conserved for \textit{all} gauges---we present in Section~\ref{sec:exoticgauge} a class of gauges for which our theory predicts that $C_{\infty}$ is finite. In Section~\ref{sec:numerics}, we verify that~$C_{\infty}$ vanishes in the Coulomb gauge with direct measurements of the relevant correlation functions in a high-resolution simulation of decaying MHD turbulence. We conclude in Section~\ref{sec:conclusion}.

\blue{\section{Sources of long-range correlations in isotropic MHD turbulence}\label{sec:invariance_perservation}\label{sec:patchsource}}


As described in Section~\ref{sec:introduction}, incompressible MHD turbulence has two non-local effects that might, in principle, correlate distant points so that $C_{\infty}\neq 0$ in~\eqref{dIHdt}. In this section, we illustrate these effects by considering a patch of non-zero velocity $\vb$ and magnetic field $\bb$ that is localised to a finite region in the vicinity of $\xb = 0$, with compact support (following~\citealt{BatchelorProudman56}). \blue{We also assume, for the sake of illustration, that we can find a vector potential $\ab$ with the same compact support.}  At the initial time, the incompressible and perfectly conducting fluid is otherwise quiescent, unmagnetised and infinite in extent. \blue{As we now explain, for any $t>0$, $\vb$ and $\ab$ do not in general have compact support---the patch at the origin generates these fields at distant points non-locally. This is the basic mechanism by which long-range correlations of the sort described in Section~\ref{sec:introduction} can in principle be established in the case of homogeneous turbulence not restricted to a compact region.}

\subsection{Far-field velocity due to the non-local pressure force\label{sec:patchpressure}}

The MHD momentum equation for an incompressible fluid is
\begin{equation}
    \frac{\p \vb}{\p t} + \vb\bcdot \bnabla \vb = - \bnabla P + \bb \bcdot \bnabla \bb+\nu \nabla^2 \vb,
\end{equation}
where $\bb$ is the magnetic field measured in units of the Alfv\'{e}n speed,  $P = (p+B^2/2)\blue{/\rho_0}$ the total pressure \blue{scaled by the constant density $\rho_0$}, $p$ the thermal pressure and $B=|\bb|$. The pressure-gradient force is determined non-locally from the condition $\bnabla \bcdot \vb = 0$, which requires
\begin{equation}
    \nabla^2 P = \blue{-}\bnabla \bcdot \left[(\vb \bcdot \bnabla) \vb - (\bb \bcdot \bnabla) \bb \right].\label{PressurePoisson}
\end{equation}
The Green's-function solution of~\eqref{PressurePoisson} is
\begin{equation}\label{eq:pressure_greens_func}
    P(\xb) = \frac{1}{4\pi} \int_{\blue{\mathbb{R}^3}} \frac{d^3 \xb^{\prime}}{| \xb^{\prime} - \xb |}\frac{\partial }{\partial x_i^{\prime}}\frac{\partial}{\partial x_j^{\prime}} \left[ u_i(\xb^{\prime} )u_j(\xb^{\prime}) -  B_i(\xb^{\prime}) B_j (\xb^{\prime})\right],
\end{equation}
which, after employing the expansion
\begin{equation}
    \frac{1}{| \xb^{\prime} - \xb |} = \frac{1}{x} - x_i^\prime \frac{\partial}{\partial x_i} \left(\frac{1}{x}\right) + \frac{1}{2} x_i^\prime x_j^\prime \frac{\partial^2}{\partial x_i \partial x_j} \left(\frac{1}{x}\right) + \dots
\end{equation}
\blue{to find the pressure field far from the origin, }becomes 
\begin{equation}\label{eqn:pressure_theoretical}
    P(\xb) = \frac{1}{4\pi} \frac{\partial^2}{\partial x_i \partial x_j}\biggl(\frac{1}{x}\biggl) \int_{\blue{\mathbb{R}^3}} d^3 \xb^{\prime} \left[ u_i(\xb^{\prime} )u_j(\xb^{\prime}) -  B_i(\xb^{\prime}) B_j (\xb^{\prime})\right] + \mathcal{O}(x^{-4})
\end{equation}
for $x=|\xb|\to \infty$ (note that the first two terms in the Taylor expansion vanish after substitution). Thus, even though $\vb$ vanishes outside the patch at the origin at $t=0$, the pressure field~\eqref{eqn:pressure_theoretical} sources it throughout space for all $t>0$.

\subsection{Far-field vector potential due to a non-local gauge \label{sec:non-localgauges}}

A second source of non-local interaction arises if $\zeta$ in~\eqref{dAdt} is a non-local function. In that case, the non-solenoidal part of $\ab$ (i.e., the part that can be written as a gradient) may become correlated between distant points. Consider\blue{, for example,} the class of gauges for which $\zeta$ is determined by the Poisson equation
\begin{equation}
    \nabla^2 \zeta = \phi,\label{poissongauge}
\end{equation}
where $\phi$ is a local function of $\vb$ and $\bb$ (i.e., depends on these fields and their spatial derivatives, but not their integrals). A prominent member of this class of gauges is the Coulomb gauge $\bnabla \bcdot \ab = 0$, for which $\phi = - \bnabla \bcdot (\vb \times \bb)$. Gauges that satisfy \eqref{poissongauge} are non-local because $\zeta$ is determined by an integral over space analogous to~\eqref{eq:pressure_greens_func}. The Green's-function solution of~\eqref{poissongauge} is
\begin{equation}\label{eq:poissongauge_greens_func}
    \zeta(\xb) = -\frac{1}{4\pi} \int_{\blue{\mathbb{R}^3}} \frac{d^3 \xb^{\prime}}{| \xb^{\prime} - \xb |}\phi(\xb').
\end{equation}
Applying~\eqref{eq:poissongauge_greens_func} for the compact patch of velocity and magnetic field considered in Section~\ref{sec:patchpressure}, we find that, for all $t>0$, a finite non-solenoidal vector potential is generated throughout space. At large $x$, the asymptotic form of $\zeta$ is
\begin{equation}\label{eqn:gauge_theoretical}
    \zeta(\xb) = -\frac{1}{4\pi x} \int_{\blue{\mathbb{R}^3}} d^3 \xb^{\prime} \phi(\xb^{\prime}) + \mathcal{O}(x^{-2}),
\end{equation}We note that, if $\phi$ is a gradient, the leading order term in~\eqref{eqn:gauge_theoretical} vanishes; for example, in the case of the Coulomb gauge, \eqref{eqn:gauge_theoretical} becomes
\begin{equation}
    \zeta_\mathrm{Coulomb}(\xb) = -\frac{1}{4\pi}\frac{\p}{\p \xb}\left(\frac{1}{x}\right) \bcdot \int_{\blue{\mathbb{R}^3}} d^3 \xb^{\prime}\, \vb(\xb^\prime) \times \bb(\xb^\prime) + \mathcal{O}(x^{-3}).\label{coulomb_zeta}
\end{equation}

\blue{Naturally, only the part of $\boldsymbol{A}$ that can be expressed as a gradient, which does not contribute to $\bb$, is affected by the non-locality~\eqref{eq:poissongauge_greens_func}---gauge effects cannot influence physical quantities. Nonetheless, this part of $\boldsymbol{A}$ does contribute to the helicity density, so contributes to $C_{\infty}$ in homogeneous turbulence.}

\blue{\section{Calculation of $C_{\infty}$ for isotropic MHD turbulence}\label{sec:calculation}}

We now turn to the problem of statistically isotropic turbulence in an infinite domain. The fact that the evolution of $\vb$ and $\ab$ at a given location in such turbulence is influenced non-locally by distant patches, as described in Section~\ref{sec:patchsource}, can lead to weak correlations between distant points that decay with separation as power laws. Let us now determine whether these effects are strong enough to enable $C_{\infty}\neq 0$~\eqref{Cinfty}, and therefore violate the conservation of $I_H$~\eqref{dIHdt}.

\subsection{\cite{BatchelorProudman56} theory of the large-$r$ asymptotics of \\correlation functions}

The theory of the asymptotic tails of correlation functions for hydrodynamic turbulence is due to \citet{BatchelorProudman56}. We review this theory briefly in this section, before applying a similar methodology to determine $C_{\infty}$ in MHD turbulence in Sections~\ref{sec:pressureinducedcorrelations} and~\ref{sec:gaugeinducedcorrelations}.

\citet{BatchelorProudman56} considered long-range correlations that arise dynamically owing to pressure-mediated interactions (Section~\ref{sec:patchpressure}) from an initial condition for which distant points are statistically independent. Stated precisely, at $t=0$, all cumulants of the velocity field decay more quickly with separation than any power law.\footnote{In brief, the cumulant is the difference between a multipoint moment (correlation function) and its decomposition into products of lower-order moments. This “connected” portion of the multi‐point moment vanishes if the points are statistically independent. \blue{See Appendix~\ref{app:cumulants} for a precise definition of the cumulant and a statement of its key properties.}} To determine correlation functions at later times, \citet{BatchelorProudman56} assumed that they could be written as convergent Taylor series in $t$, with derivatives evaluated at $t=0$. The terms in the Taylor series that decay most slowly with the separation $r$ give the large-$r$ asymptotic of the correlation function at $t>0$. 


To illustrate the method, consider the velocity triple correlation $\langle u_i u_j u_k^\prime \rangle$, where primed variables are evaluated at $\xb^{\prime} = \xb + \rb$ and unprimed variables are evaluated at $\xb$. Its first time derivative involves the non-locally determined pressure:
\begin{align}
    \frac{\p}{\p t}\langle u_i u_j u_k' \rangle & = \dots - \langle u_i u_j \frac{\p P^\prime}{\p x_k^\prime} \rangle + \dots\label{dt_triplecorrelator1}
    \\ & = \dots - \frac{\p}{\p r_k}\langle u_i u_j  \frac{1}{4\pi} \int_{\blue{\mathbb{R}^3}} \frac{d^3 \xb^{\prime \prime}}{| \xb^{\prime \prime} - \xb ^\prime|}\frac{\partial }{\partial x_l^{\prime \prime}}\frac{\partial}{\partial x_m^{\prime \prime}} u_l^{\prime\prime} u_m^{\prime\prime}\rangle + \dots \label{dt_triplecorrelator}
\end{align}
where we have used~\eqref{eq:pressure_greens_func} (dropping the terms involving the magnetic field, which are not present in \citeauthor{BatchelorProudman56}'s analysis, but behave in the same way as the velocity terms and do not change the result). We define $\Sb = \xb - \xb^{\prime \prime}$, so that
\begin{equation}
    \frac{1}{|\xb^{\prime \prime}-\xb^{\prime}|}=\frac{1}{|\rb+\Sb|}= \frac{1}{r}+s_i\frac{\p}{\p r_i}\frac{1}{r}+\blue{\frac{1}{2}}s_i s_j \frac{\p}{\p r_i}\frac{\p}{\p r_j}\frac{1}{r}+\mathcal{O}\left(\frac{1}{r^4}\right).\label{s_expansion}
\end{equation}
Substituting~\eqref{s_expansion} into~\eqref{dt_triplecorrelator} yields
\begin{equation}
    \frac{\p}{\p t}\langle u_i u_j u_k' \rangle = \dots - \frac{1}{4\pi} \frac{\p^3}{\p r_k \partial r_l \partial r_m} \left(\frac{1}{r}\right)\int_{\blue{\mathbb{R}^3}} d^3 \Sb \left(\langle u_i u_j u_l^{\prime \prime}  u_m^{\prime \prime}\rangle  - \langle u_i u_j \rangle\langle u_l^{\prime \prime}  u_m^{\prime \prime} \rangle\right)\dots,\label{triplecorrelator}
\end{equation}
where, as in~\eqref{eqn:pressure_theoretical}, the first two terms in the series~\eqref{s_expansion} vanish. We now evaluate~\eqref{triplecorrelator} at $t=0$, at which time, the decay of all cumulants being faster than any power law, all their integral moments converge, so all the higher order terms arising from the expansion~\eqref{s_expansion} converge. No stronger dependence on $r$ exists in any term in the Taylor expansion in $t$, hence
\begin{equation}
    \langle u_i u_j u_k' \rangle = \mathcal{O}\left(\frac{1}{r^4}\right), \quad t>0.\label{hydrotriplecorrelator}
\end{equation}\blue{We note that the $r^{-4}$ scaling of the velocity triple-correlation function~\eqref{hydrotriplecorrelator} is responsible for the non-conservation of the \cite{Loitsyansky39} integral in decaying hydrodynamic turbulence---see~\cite{Davidson15} for a review.}

\subsection{The case of a local gauge function~$\zeta$\label{sec:pressureinducedcorrelations}}

We now apply a similar analysis to the one that led to~\eqref{hydrotriplecorrelator} to the problem of determining $C_{\infty}$ [Equation~\eqref{Cinfty}]. We shall take the statistical independence of distant points at $t=0$ to mean that all cumulants involving $\vb$, $\ab$ and $\bb$ decay with separation faster than any power law at $t=0$. Under this assumption, we seek the large-$r$ tails of the correlation functions appearing in~\eqref{Cinfty}. \blue{We do not make any assumption about the relative sizes of the fields $\vb$ and $\bb$ (i.e., about the Alfv\'enic Mach number) in the initial state.}\footnote{\blue{We do assume that the resistive terms in~\eqref{dIHdt} remain negligible, which, in practice, means that the velocity field is not strong enough to shear the integral-scale magnetic field to resistive scales.}} We shall first consider a local gauge---i.e., one for which $\zeta$ is a local function of $\ab$, $\bb$ and $\vb$---and examine pressure-induced correlations (Section~\ref{sec:patchpressure}) only, returning to the issue of gauge-induced correlations (Section~\ref{sec:non-localgauges}) in Section~\ref{sec:gaugeinducedcorrelations}.  We shall find that all terms \blue{in the Taylor expansion in which the total pressure $P$ appears exactly once, which might lead to the correlation functions decay as $r^{-4}$, as in~\eqref{dt_triplecorrelator1}, vanish due to symmetry considerations. This means that the only surviving terms are ones in which $P$ appears more than once, which decay with separation as $r^{-8}$ or faster. Thus, $C_{\infty}=0$. We first present a specific example of a contribution that vanishes, then give a general argument that all similar contributions also vanish.}

\subsubsection{Explicit example of a vanishing term}

 The two fifth-order correlation functions that appear in~\eqref{Cinfty} \blue{for $C_{\infty}$} have general form $\langle u_iA_jB_kA_l'B_l'\rangle$, with contractions over different pairs of the free indices $ijk$. \blue{Let us consider their Taylor expansion in time, evaluated at $t=0$. We first present a specific example of a contribution to the Taylor series that vanishes, then give a general argument that all similar contributions (i.e., those with one appearance of $P$) also vanishes.}
 
 Differentiating $\langle u_iA_jB_kA_l'B_l'\rangle$ twice with respect to time yields terms involving the non-locally determined pressure:
\begin{align}
    &\quad \frac{\p^2 }{\p t^2}\langle u_iA_jB_kA_l'B_l'\rangle \nonumber\\ &
    =\dots - \bigg\langle \frac{\p u_m}{\p t} \frac{\p u_i}{\p x_m}A_jB_kA_l'B_l'\bigg\rangle \dots \nonumber\\ &
    = \dots +\bigg\langle \left(\frac{\p}{\p x_m}\frac{1}{4\pi} \int_{\blue{\mathbb{R}^3}} \frac{d^3 \xb^{\prime\prime}}{| \xb^{\prime\prime} - \xb |}\frac{\partial }{\partial x_r^{\prime\prime}}\frac{\partial}{\partial x_s^{\prime\prime}} \left( u_r^{\prime\prime}u_s^{\prime\prime} -  B_r^{\prime\prime} B_s^{\prime\prime}\right)\right) \frac{\p u_i}{\p x_m} A_j B_k A^{\prime}_l B^\prime_l\bigg\rangle \dots \nonumber
    \\ &
    = \dots +\frac{\p}{\p r_m}\frac{1}{4\pi} \int_{\blue{\mathbb{R}^3}} \frac{d^3 \xb^{\prime\prime}}{| \xb^{\prime\prime} - \xb |}\frac{\partial }{\partial x_r^{\prime\prime}}\frac{\partial}{\partial x_s^{\prime\prime}}\bigg\langle B_r^{\prime\prime} B_s^{\prime\prime} \frac{\p u_i}{\p x_m} A_j B_k A^{\prime}_l B^\prime_l\bigg\rangle 
    \nonumber\\ &\quad \,\,\,\quad + \frac{1}{4\pi} \int_{\blue{\mathbb{R}^3}} \frac{d^3 \xb^{\prime\prime}}{| \xb^{\prime\prime} - \xb |}\frac{\partial }{\partial x_r^{\prime\prime}}\frac{\partial}{\partial x_s^{\prime\prime}} \bigg\langle B_r^{\prime\prime} B_s^{\prime\prime} \frac{\p}{\p x_m}\left( \frac{\p u_i}{\p x_m} A_j B_k\right) A^{\prime}_l B^\prime_l\bigg\rangle+\dots,\label{dt3_5thcorrelator}
\end{align}where in the \blue{second equality we have substituted~\eqref{eq:pressure_greens_func}, and in the} third equality we have restricted attention to the terms inside the integral that involve $\bb$, the ones that involve~$\vb$ being analogous. 

We now evaluate~\eqref{dt3_5thcorrelator} at the initial time in the limit of large $r=|\xb^\prime-\xb|$, \blue{seeking a result analogous to~\eqref{triplecorrelator}}. Unlike in~\eqref{dt_triplecorrelator}, the correlation functions in the final line of \eqref{dt3_5thcorrelator} \blue{contain fields} evaluated at three different points, viz., $\xb''$, $\xb'$ and $\xb$. \blue{Let us define $\Sb=\xb''-\xb$ and $\Sb'=\xb''-\xb'$ to be the two independent displacement vectors between these points. Because $\rb = \xb^\prime - \xb = \Sb - \Sb^\prime$, it must be the case that $|\Sb|\sim r$ or $ |\Sb'|\sim r$ (or both) as $r\to\infty$. In the former case, \blue{our assumption about the vanishing of (second-order) cumulants (see start of Section~\ref{sec:pressureinducedcorrelations}) means that} correlation functions of the form $\langle X Y^\prime Z^{\prime\prime}\rangle\to\langle X \rangle \langle Y^\prime Z^{\prime\prime}\rangle + o(r^{-n})$ for all $n>0$ as $r\to \infty$, while in the latter case, $\langle X Y^\prime Z^{\prime\prime}\rangle\to\langle Y^\prime \rangle \langle X Z^{\prime\prime}\rangle+ o(r^{-n})$. Applied to the first of the two integrals appearing after the final equality in~\eqref{dt3_5thcorrelator}, these formulae become}
\begin{equation}
\bigg\langle B_r^{\prime\prime} B_s^{\prime\prime} \frac{\p u_i}{\p x_m} A_j B_k A^{\prime}_l B^\prime_l\bigg\rangle \to \begin{cases}
    \displaystyle\bigg\langle B_r^{\prime\prime} B_s^{\prime\prime} \frac{\p u_i}{\p x_m} A_j B_k \bigg\rangle \langle A^{\prime}_l B^\prime_l\rangle + o(r^{-n})& \text{\blue{if $|\Sb'|\sim r$},} \vspace{2mm}\\
    \displaystyle\bigg\langle  \frac{\p u_i}{\p x_m} A_j B_k \bigg\rangle \langle B_r^{\prime\prime} B_s^{\prime\prime} A^{\prime}_l B^\prime_l\rangle + o(r^{-n}) & \text{\blue{if $|\Sb|\sim r$},}
\end{cases}\label{cases_pressure_correlations}
\end{equation}\blue{for all $n>0$ as $r\to\infty$. In both cases, the leading-order term vanishes. In the first case, this is because $\langle A_l' B_l'\rangle = 0$. In the second case, the correlation function} $\langle B_r^{\prime\prime} B_s^{\prime\prime} A^{\prime}_l B^\prime_l\rangle$ must change sign under reflection, because $\bb$ is an axial (pseudo-) vector and $\ab$ is a polar (true) vector.\footnote{\blue{For completeness, we note that, in principle, it is possible that $\ab$ is not a polar vector---for any polar vector $\ab$, we can create a new vector potential of mixed parity by adding the gradient of a pseudo-scalar.}} In reflection-symmetric turbulence, the most general decomposition of this correlation function consistent with this fact is \blue{(see, e.g., \citealt{robertson1940invariant} or~\citealt{batchelor1953THT})} 
\begin{equation}
    \langle B_r^{\prime\prime} B_s^{\prime\prime} A^{\prime}_l B^\prime_l\rangle = a(|\Sb^\prime|) \epsilon_{rsn} s_n^\prime,
\end{equation}for which $a(|\Sb^\prime|)$ must be zero given that the left-hand side is symmetric under exchanging the indices $r$ and $s$.

In the equivalent expression \blue{to~\eqref{cases_pressure_correlations} for the second of the two integrals appearing after the final equality in~\eqref{dt3_5thcorrelator}}, both cases are zero: $\langle A_l' B_l'\rangle = 0$ by assumption and $\langle \p_m[(\p_m u_i)A_jB_k]\rangle = 0$ \blue{because $\langle \p_m (\dots)\rangle = 0$ for homogeneous turbulence}. 

\subsubsection{Generalisation to other terms in the Taylor expansion}

Let us now consider whether there are \blue{\emph{any} terms in the Taylor expansion of $\langle u_iA_jB_kA_l'B_l'\rangle$ that do not vanish after contraction of $i$ with $j$ or of $j$ with $k$, as in $C_{\infty}$~\eqref{Cinfty}.}
Consider terms in which the total pressure $P$ appears once. Such terms have the general form
\begin{equation}
    C_{i_1 i_2 \dots i_n} \frac{\p^{m}}{\p r_{j_1} \p r_{j_2} \dots\p r_{j_m}} \left(\frac{1}{r}\right)\int_{\blue{\mathbb{R}^3}} d^3 \Sb \, D_{k_1 k_2 \dots k_p}(\Sb)\label{1integral}
\end{equation}where $C_{i_1 i_2 \dots i_n}$ and $D_{k_1 k_2 \dots k_m}$ are pseudotensors.  \blue{Since the quantity~\eqref{1integral} must be a vector, all but one of its indices are contracted. However, none of the indices $i_1 \dots i_n$ can be contracted with the indices $k_1 \dots k_m$, because independent fields (for which the correlation functions can be split) cannot appear in contraction in the Taylor expansion.} Thus, all of these indices, with the possible exception of one (the single free index) must be contracted with the indices $j_1 \dots j_m$. 
However, the indices $j_1 \dots j_m$ are symmetric under interchange, while $C_{i_1 i_2 \dots i_n}$ and $D_{k_1 k_2 \dots k_m}$ are each antisymmetric under the exchange of at least two of their indices, by virtue of being pseudotensors. \blue{It follows that}, all terms having the form~\eqref{1integral}---i.e., having one appearance of pressure---are zero. 

\blue{We conclude} that the only terms in the Taylor expansion that are non-vanishing are the ones for which pressure appears twice. \blue{Because these involve two appearances of $\p_i P$, they decay as $r^{-8}$ or faster, so we have that} 
\begin{equation}
    \langle u_iA_jB_kA_l'B_l'\rangle = \mathcal{O}\left(\frac{1}{r^8}\right).
\end{equation}Precisely analogous arguments to those just given imply that $\langle \zeta B_iA_l'B_l'\rangle = \mathcal{O}(r^{-8})$.
\footnote{We do not exclude the possibility that all power-law contributions to the correlation functions vanish, including those for which pressure appears more than once. However, we have not been able to find a general proof of this.}

Because the tails of all the correlation functions that appear in~\eqref{Cinfty} decay faster than~$r^{-2}$ as $r\to\infty$, we conclude that $C_{\infty}=0$ if $\zeta$ is a local function.

\subsection{The case of non-local $\zeta$ \label{sec:gaugeinducedcorrelations}}

We turn now to long-range interactions arising from the use of a non-local gauge function~$\zeta$, which can cause the non-solenoidal part of $\ab$ to become correlated between distant points, as explained in Section~\ref{sec:patchsource}. 

\subsubsection{Poisson-type non-locality}

We prove in this section that sufficiently strong correlations for $C_{\infty}\neq 0$ never arise for any gauge for which $\zeta$ is determined by Poisson's equation~\eqref{poissongauge}, with the source $\phi$ a local function of $\vb$ and $\bb$. For such gauges, we can evaluate all the relevant contributions to $C_{\infty}$ explicitly: because $\zeta$ only appears in terms for which $\ab$ is differentiated with respect to time, the number of such terms is limited. We address them in turn below.

As we utilised in Section~\ref{sec:pressureinducedcorrelations}, the first two correlation functions in~\eqref{Cinfty} have general form $\langle u_iA_jB_kA_l'B_l'\rangle$. Differentiating this once with respect to time and isolating the gauge terms, we have
\begin{align}
\frac{\partial}{\partial t}\langle u_iA_jB_kA_l'B_l'\rangle 
&= \dots \bigg\langle u_i \frac{\partial \zeta}{\partial x_j} B_kA_l'B_l'\bigg\rangle
   + \frac{\partial}{\partial r_l}\langle u_iA_jB_k \zeta' B_l'\rangle
   + \dots
   \nonumber \\
& = \dots
    \frac{\partial}{\partial r_j}\frac{1}{4\pi}
     \int_{\blue{\mathbb{R}^3}} \frac{d^3\xb''}{|\xb'' - \xb|}
     \langle u_i \phi''B_kA_l'B_l'\rangle
   \nonumber \\
& \phantom{= \dots }
   + \frac{1}{4\pi}
     \int_{\blue{\mathbb{R}^3}} \frac{d^3\xb''}{|\xb'' - \xb|}
     \bigg\langle \frac{\partial}{\partial x_j}(u_iB_k)\phi''A_l'B_l'\bigg\rangle
   \nonumber \\
& \phantom{= \dots}
   - \frac{\partial}{\partial r_l}\,\frac{1}{4\pi}
     \int_{\blue{\mathbb{R}^3}} \frac{d^3\xb''}{|\xb'' - \xb'|}
     \langle u_iA_jB_k \phi'' B_l'\rangle
   + \dots\label{gauge1stderiv}
\end{align}Taking the limit $r=|\xb' - \xb|\to\infty$, the correlation functions appearing inside the three integrals in~\eqref{gauge1stderiv} split, as in~\eqref{cases_pressure_correlations}. In the first integral,
\begin{equation}
\langle u_i \phi''B_kA_l'B_l'\rangle \to 
\begin{cases}
    \langle u_i \phi''B_k\rangle \langle A_l'B_l'\rangle + o(r^{-n}) & \text{\blue{if $|\Sb'|\sim r$},} \vspace{2mm}\\
    \langle u_i B_k \rangle \langle \phi'' A_l'B_l'\rangle + o(r^{-n}) & \text{\blue{if $|\Sb|\sim r$}.}
\end{cases}\label{integral1}
\end{equation}
In the second, 
\begin{equation}
\langle \p_j(u_iB_k)\phi''A_l'B_l'\rangle \to 
\begin{cases}
    \langle \p_j(u_iB_k)\phi''\rangle \langle A_l'B_l'\rangle + o(r^{-n}) & \text{\blue{if $|\Sb'|\sim r$},} \vspace{2mm}\\
    \langle \p_j(u_iB_k)\rangle \langle \phi''A_l'B_l'\rangle + o(r^{-n}) & \text{\blue{if $|\Sb|\sim r$}.}
\end{cases}\label{integral2}
\end{equation}
In the third, 
\begin{equation}
\langle u_iA_jB_k \phi'' B_l'\rangle \to 
\begin{cases}
    \langle u_iA_jB_k \phi'' \rangle \langle B_l'\rangle + o(r^{-n}) & \text{\blue{if $|\Sb'|\sim r$},} \vspace{2mm}\\
    \langle u_iA_jB_k \rangle \langle \phi'' B_l'\rangle + o(r^{-n}) & \text{\blue{if $|\Sb|\sim r$}.}
\end{cases}\label{integral3}
\end{equation}In all of these cases, the product of the split correlation function is zero, as follows. In \eqref{integral1} and \eqref{integral2}, the cases with \blue{$|\Sb'|\sim r$} vanish because $\langle A_l B_l\rangle = 0$ in reflection-symmetric turbulence. In \eqref{integral3}, the case with \blue{$|\Sb'|\sim r$} finite vanishes because $\langle B_l\rangle = 0$, by isotropy. Of the cases for which \blue{$|\Sb|\sim r$}: in \eqref{integral1}, $\langle u_i B_k \rangle = \delta_{ij}\langle \vb \bcdot \bb\rangle/3=0$ in reflection-symmetric turbulence; in \eqref{integral2}, $\langle \p_j(u_iB_k)\rangle=0$ in homogeneous turbulence; and in \eqref{integral3}, $\langle \phi'' B_l'\rangle = 0$ in homogeneous turbulence because $B_l$ is solenoidal. It follows that there is no contribution to $\p_t\langle u_iA_jB_kA_l'B_l'\rangle$ from the gauge at $t=0$.

Let us now consider the second time derivative of $\langle u_iA_jB_kA_l'B_l'\rangle$. Its only non-trivial term that is qualitatively distinct from those appearing in~\eqref{gauge1stderiv} is the one for which the two appearances of the vector potential are each differentiated with respect to time:
\begin{align}
\frac{\p^2}{\p \blue{t^2} }\langle u_iA_jB_kA_l'B_l'\rangle 
&= \dots \frac{\partial}{\partial r_l}\langle u_i \frac{\partial \zeta}{\partial x_j} B_k \zeta' B_l'\rangle
   + \dots
   \nonumber \\
& = \dots
   - \frac{\partial^2}{\partial r_l r_j}\frac{1}{(4\pi)^2}
     \int_{\blue{\mathbb{R}^3}} \frac{d^3\xb''}{|\xb'' - \xb|}\int_{\blue{\mathbb{R}^3}} \frac{d^3\xb'''}{|\xb''' - \xb'|}
     \langle u_i \phi''B_k\phi'''B_l'\rangle
   \nonumber \\
& \phantom{= \dots }
   - \frac{\p}{\p r_l}\frac{1}{(4\pi)^2}
     \int_{\blue{\mathbb{R}^3}} \frac{d^3\xb''}{|\xb'' - \xb|}\int_{\blue{\mathbb{R}^3}} \frac{d^3\xb'''}{|\xb''' - \xb'|}
     \bigg\langle \frac{\partial}{\partial x_j}(u_iB_k)\phi''\phi '''B_l'\bigg\rangle+\dots.\label{dt2_5thcorrelator_gauge}
\end{align}\blue{The correlation functions appearing in the final line of \eqref{dt2_5thcorrelator_gauge} involve fields evaluated at four points, $\xb$, $\xb^\prime$, $\xb^{\prime \prime}$ and $\xb^{\prime \prime \prime}$. In the limit $r\to\infty$, there are four qualitatively distinct possibilities for how these points can be arranged. The first is that all of their separations are $\sim r$, in which case the correlation functions vanish with $r$ faster than any power law at $t=0$. The remaining three cases involve some of the separations being held finite as the limit is taken. As concerns the correlation function in the first integral, the possibilities are}
\begin{equation}
\langle u_i \phi''B_k\phi'''B_l'\rangle \to 
\begin{cases}
    \langle u_i \phi''B_k\phi'''\rangle\langle B_l'\rangle + o(r^{-n}) & \text{for $|\xb^{\prime \prime}-\xb|$, $|\xb^{\prime \prime \prime}-\xb|$ finite,}\\
    \langle u_i B_k\phi'''\rangle \langle\phi''B_l'\rangle + o(r^{-n}) & \text{for $|\xb^{\prime \prime}-\xb^{\prime}|$, $|\xb^{\prime \prime \prime}-\xb|$ finite.}\\
    \langle u_i B_k\rangle \langle \phi''\phi'''B_l'\rangle + o(r^{-n}) & \text{for $|\xb^{\prime \prime}-\xb^{\prime}|$, $|\xb^{\prime \prime \prime}-\xb^\prime|$ finite.}
\end{cases}\label{eq:double_cases}
\end{equation}\blue{Because $\langle B_l'\rangle$, $\langle \phi'' B_l'\rangle$, and $ \langle u_i B_k \rangle$ are all zero, each of the cases in~\eqref{eq:double_cases} are zero}. As concerns the second line of~\eqref{dt2_5thcorrelator_gauge}, it is readily verified that replacing $u_i B_k$ in~\eqref{eq:double_cases} by $\p_j(u_i B_k)$ yields terms that still vanish. We conclude that the terms presented explicitly in \eqref{dt2_5thcorrelator_gauge} vanish. By precisely analogous reasoning, there are also no terms in the Taylor expansion in time of $\langle \zeta u_iA_l'B_l'\rangle$ that have power-law tails in $r$ induced by substituting~\eqref{eq:poissongauge_greens_func}. 

In the above analysis, we have not included terms for which pressure and gauge enter together---such terms decay more quickly with $r$ than would pressure-only terms, i.e., as $\mathcal{O}(r^{-5})$\blue{, so cannot lead to $C_{\infty}\neq 0$}. We have also not treated the case where $\phi$ is a function of $\ab$ as well as of $\bb$ and $\vb$; we anticipate that, for a large class of such gauges, $C_{\infty}$ does vanish, but we have not proven this in general. 

\subsubsection{An $I_H$-non-conserving gauge\label{sec:exoticgauge}}

We have so far shown that $C_{\infty}$ vanishes under a wide class of gauge choices, including all those for which $\zeta$ is determined by Poisson's equation~\eqref{poissongauge}, with $\phi$ a local function of $\vb$ and $\bb$. Let us now ask whether there exist more exotic gauges for which $C_{\infty}\neq 0$. Interestingly, the answer appears to be yes. An explicit example is the gauge defined by
\begin{equation}
    \zeta = B_i \mathcal{Z}_i,\quad \nabla^2 \mathcal{Z}_i = \phi B_i,\label{funkygauge}
\end{equation}where $\phi$ is any local function of $\vb$, $\bb$ and $\ab$. The Green's-function solution for $\zeta$ is
\begin{equation}\label{eq:funkygauge_greens_func}
    \zeta(\xb) = -\frac{B_i}{4\pi} \int_{\blue{\mathbb{R}^3}} \frac{d^3 \xb^{\prime}}{| \xb^{\prime} - \xb |}B_i'\phi'.
\end{equation}It follows that, at $t=0$, the third correlation function \blue{that appears in the definition of $C_{\infty}$}~\eqref{Cinfty} is
\begin{align}
    \langle \zeta B_iA_l'B_l'\rangle & = - \frac{1}{4\pi}\int_{\blue{\mathbb{R}^3}} \frac{d^3 \xb^{\prime\prime}}{| \xb^{\prime\prime} - \xb |}\langle   B_k B_i B_k''\phi'' A_l'B_l'\rangle\nonumber\\
    & = -\frac{1}{4\pi}\langle   B_k B_i \rangle\int_{\blue{\mathbb{R}^3}} \frac{d^3 \xb^{\prime\prime}}{| \xb^{\prime\prime} - \xb |} \langle B_k''\phi'' A_l'B_l'\rangle\nonumber\\
    & = -\frac{\langle   B^2 \rangle}{12\pi}\int_{\blue{\mathbb{R}^3}} \frac{d^3 \Sb^\prime}{| \Sb^\prime + \rb |} \langle B_i''\phi'' A_l'B_l'\rangle\nonumber\\
    & = -\frac{\langle   B^2 \rangle}{12\pi} \frac{1}{r}\int_{\blue{\mathbb{R}^3}} d^3 \Sb^\prime \langle B_i''\phi'' A_l'B_l'\rangle- \frac{\langle   B^2 \rangle}{12\pi} \frac{\p}{\p r_j}\left(\frac{1}{r}\right)\int_{\blue{\mathbb{R}^3}} d^3 \Sb^\prime s_j^\prime \langle B_i''\phi'' A_l'B_l'\rangle + \mathcal{O}(r^{-3}).\label{exoticgaugeevolution}
\end{align}Provided that $\phi$ is not a true scalar (the natural choice would, presumably, be the pseudo-scalar $\phi~\blue{\propto}~h = \ab\bcdot\bb$), the correlation function appearing in the above is
\begin{equation}
    \langle B_i''\phi'' A_l'B_l'\rangle = a(|\Sb^\prime|)s_i^\prime,
\end{equation}where $a(|\Sb^\prime|)$ is an undetermined function. This does not, in general, vanish, although the first integral in~\eqref{exoticgaugeevolution} vanishes by symmetry. The second integral in~\eqref{exoticgaugeevolution} does not obviously vanish, so we expect that ${\langle \zeta B_iA_l'B_l'\rangle = \mathcal{O}(r^{-2})}$, indicating that $C_{\infty}$ is finite for this gauge [see~\eqref{Cinfty}]. We suggest a possible interpretation of this phenomenon in Section~\ref{sec:conclusion}, but defer detailed investigation of this interesting gauge to future work.

\section{Numerical results\label{sec:numerics}}

In this section, we present measurements of the three correlation functions that contribute to $C_{\infty}$~\eqref{Cinfty} in a high-resolution direct numerical simulation (DNS) of decaying MHD turbulence using the Coulomb gauge for $\ab$. We aim to establish whether the {large-$r$} tails of these correlation functions indeed decay sufficiently quickly to ensure the conservation of $I_H$ (i.e., faster than $r^{-2}$), as predicted in Section~\ref{sec:calculation}. \blue{We first describe our numerical method and initial conditions in Section~\ref{sec:initconds}, then the features of the evolution in Section~\ref{sec:overall}. We discuss our measurements of correlation functions in Section~\ref{sec:measurements}. }

\subsection{Numerical method and initial conditions\label{sec:initconds}}
We employ a modified version of the \textsc{flash} code \citep{FryxellEtAl2000,DubeyEtAl2008,federrath2021sonic} with the 5-wave HLL5R approximate Riemann solver~\citep{WaaganFederrathKlingenberg2011}, which uses a positivity-preserving MUSCL-Hancock scheme \citep{bouchut2007multiwave,bouchut2010multiwave,waagan2009positive} to solve the equations of three-dimensional compressible isothermal MHD\blue{---i.e.,
\begin{equation}
\rho\left(\frac{\p \vb}{\p t} + \vb\bcdot\bnabla \vb\right) = -\bnabla p + \frac{1}{4\pi}(\nabla\times\bb)\times\bb+\rho \nu \nabla^2 \vb,
\end{equation}
\begin{equation}
\frac{\p \bb}{\p t} = \nabla\times(\vb\times\bb) + \eta\nabla^2 \bb,
\end{equation}and $p = \rho c_s^2$, with $c_s$ the constant speed of sound ($c_s = 1$ in our code units)---}in a periodic box of volume $L^3$\blue{, where $L=1$ in our code units. Because our simulations evolve a compressible fluid, we shall in what follows measure the magnetic field $\bb$ in full CGS (Gaussian) units, rather than in units of the Alfv\'{e}n speed, as was convenient in the preceding sections.} We use a uniform grid with $2304^3$ cells (see Appendix~\ref{sec:resolution_study} for comparison with a simulation at lower resolution). The simulation uses explicit kinematic viscosity ${\nu=10^{-7}\,L\cs}$ and magnetic diffusivity $\eta=10^{-7}\,L\cs$, so the magnetic Prandtl number is ${\prandtl\equiv\nu/\eta=1}$.

The simulation is initialised with a uniform density $\rho_0$\blue{, with $\rho_0=1$ in our code units,} and zero velocity.\footnotemark~As in \citet{HoskingSchekochihin20decay} and \citet{Zhou22_Hosking}, the initial condition for the magnetic field is a non-helical Gaussian random field.  We choose a magnetic-energy spectrum ${\emag(k)\propto k^4}$ for $1\leq kL/2\pi \leq60 \equiv k_{\mathrm{peak},0} L /2\pi$, and zero elsewhere, generated with \texttt{TurbGen} \citep{FederrathDuvalKlessenSchmidtMacLow2010,FederrathEtAl2022ascl}. \blue{We choose the initial amplitude of the magnetic field such that $\langle B_0^2\rangle = \rho_0 c_s^2$, i.e., the initial (root-mean-square) Alfv\'{e}n speed is $v_{A0}\equiv \langle B_0^2\rangle^{1/2}/\sqrt{4\pi\rho_0} = c_s/\sqrt{4\pi}\simeq 0.3c_s$. The mean magnetic field in the periodic domain (i.e., its $\bk=0$ component) is zero. Thus, the magnetic vector potential is well-defined, as is the net magnetic helicity in the periodic box, which is equal to zero.} Using~$\emag$ we define the magnetic-field correlation scale (integral scale) by
\begin{equation} \label{xi_M}
    \lmag = \blue{\frac{2\pi}{E_M}}\int_0^{\infty} dk\,\frac{\emag(k)}{k}
\end{equation}where $E_M = \int_0^{\infty} dk\,\emag(k)$ is the magnetic energy.
The initial Lundquist number is ${\lundquist_0 \equiv \va \lmagi/\eta \simeq 6\times10^4}$, where $\lmagi=\blue{2\pi}~\blue{\times }~5/(4k_{\mathrm{peak},0})$ is the initial $\lmag$. \blue{In the results presented below, we restrict attention to times for which $\xi_M \ll L$; as far as any individual magnetic structure is concerned, box-scale topological effects~\citep{Berger97} can then be neglected, and the periodic box serves as a proxy for an infinite open domain, as considered in Sections~\ref{sec:invariance_perservation} and~\ref{sec:calculation}.}

\footnotetext{\blue{The calculation of $C_{\infty}$ in Section~\ref{sec:calculation} did not assume zero initial velocity; we choose the magnetically dominated initial condition here as a clean numerical experiment. We note that the decay laws~\eqref{nonhelicallaws} can be different when the initial velocity is non-zero, because of constraints from other dynamical invariants; see~\cite{HoskingSchekochihin20decay} and references therein.}}



\subsection{\blue{Evolution of the simulation}\label{sec:overall}}

\begin{figure}
\centering
\includegraphics[width = 1\linewidth]{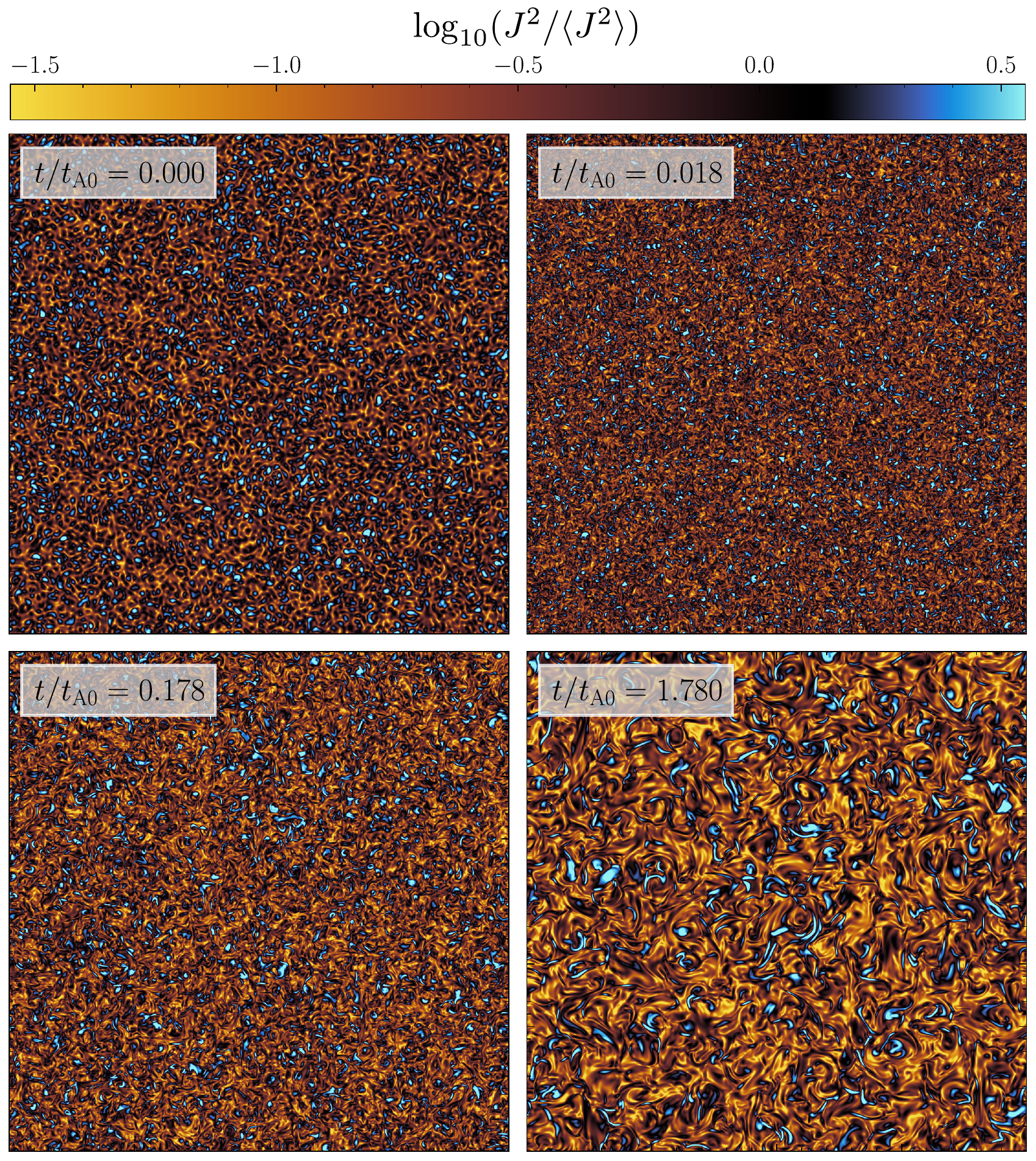}
\caption{\label{fig:current_density} Two-dimensional slices of the current density squared $J^2 = |\bnabla \times \bb|^2$, normalised to its root-mean-square value, at various times during the decay. After $t \sim t_{A0}$, we see growth of magnetic structures resulting from the merging of smaller structures.}
\end{figure}

The out-of-equilibrium initial condition (see Section~\ref{sec:initconds}) undergoes turbulent decay as visualised in figure~\ref{fig:current_density}, which shows slices of the squared current density at different times. We observe that the coherence scale of the magnetic field increases with time, as expected from the decay laws~\eqref{nonhelicallaws}.

\begin{figure}
\begin{subfigure}{0.48\textwidth}
\centering
\includegraphics[ width = 1\linewidth]{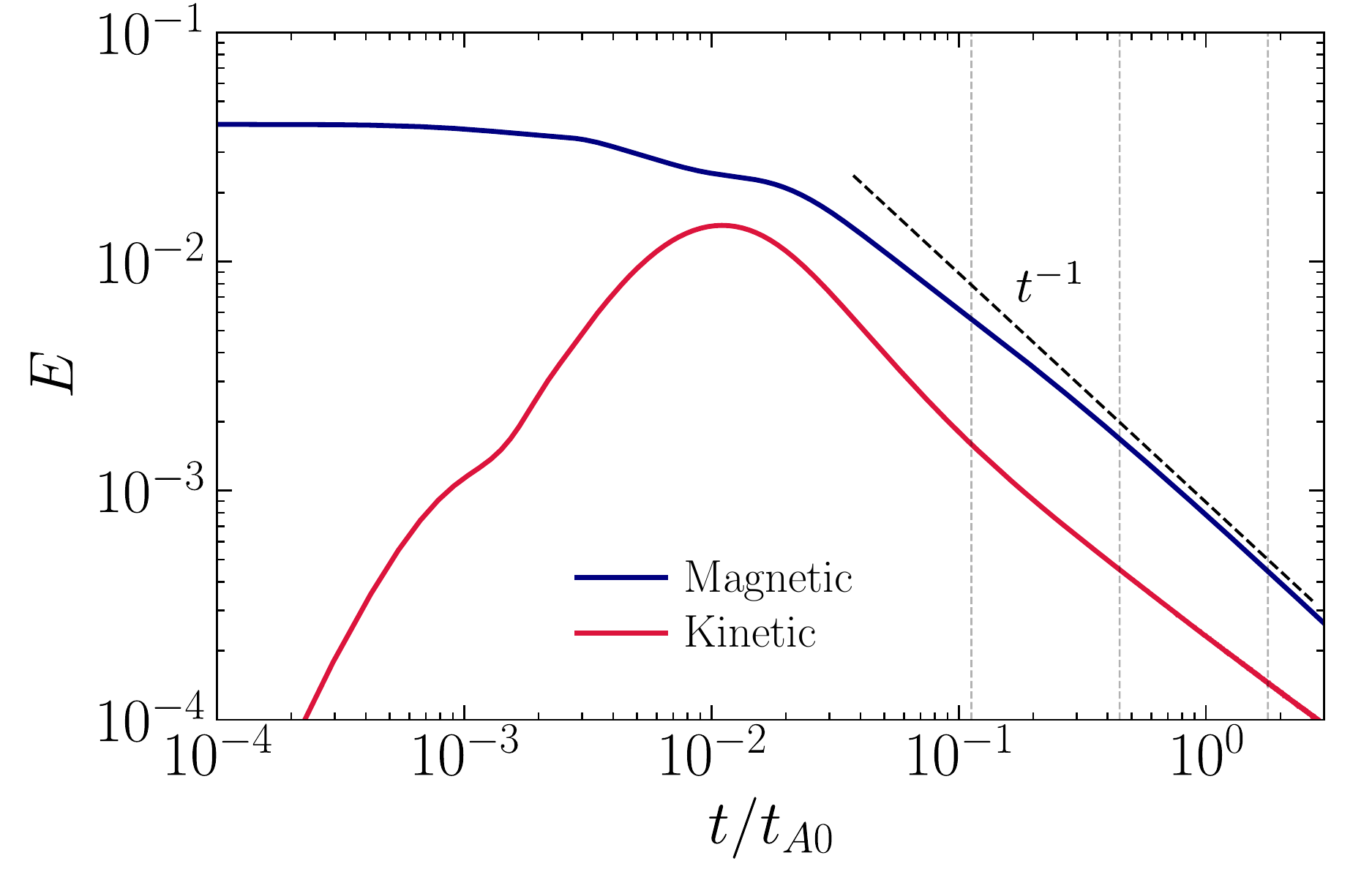}
\caption{\label{fig:ekinemag}} 
\end{subfigure}
\begin{subfigure}{0.48
\textwidth}
\centering
\includegraphics[width = 1\linewidth]{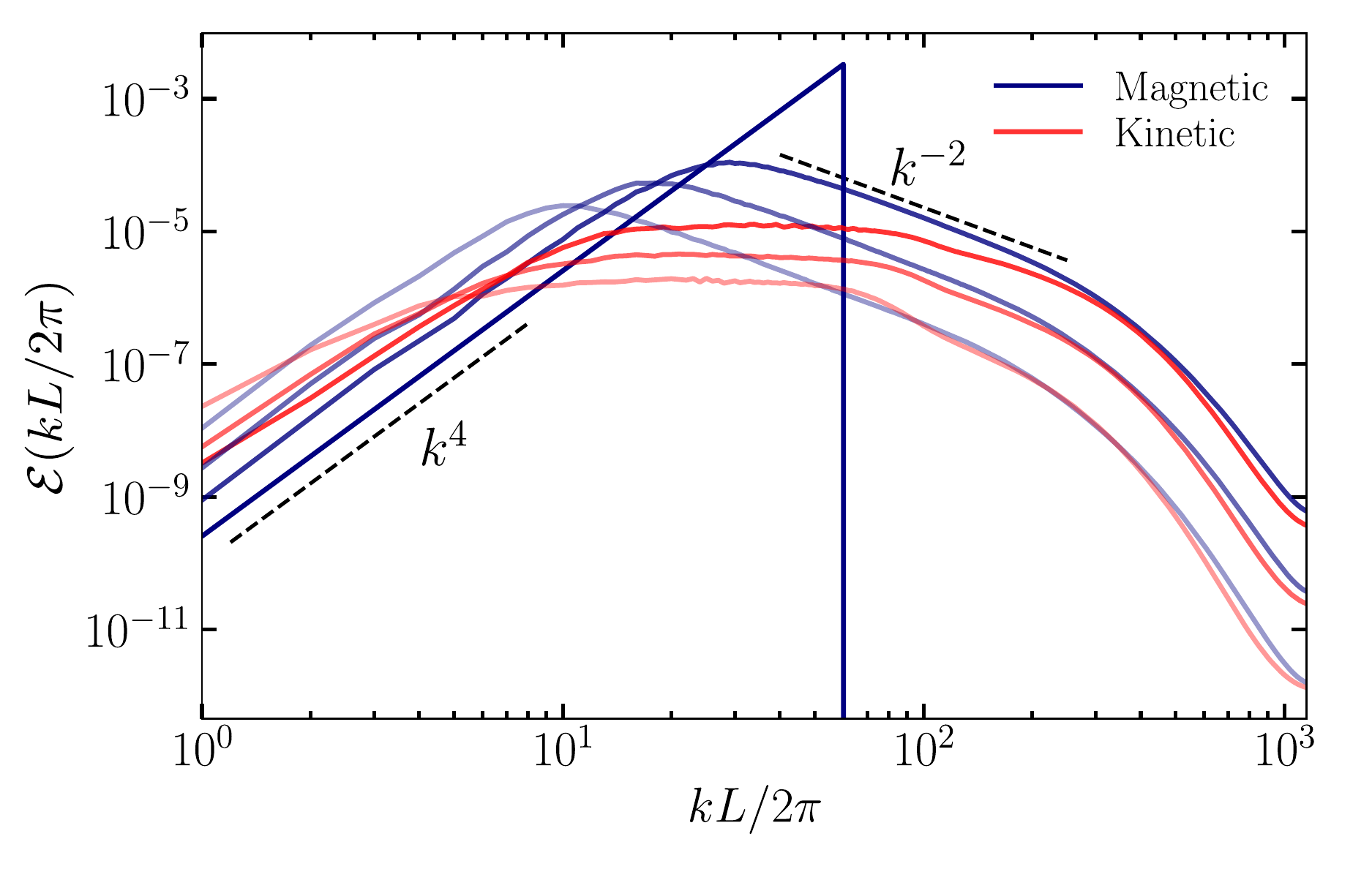}
\caption{\label{fig:spectra} }
\end{subfigure}
\begin{subfigure}{
\textwidth}
\centering
\includegraphics[width = 0.6\linewidth]{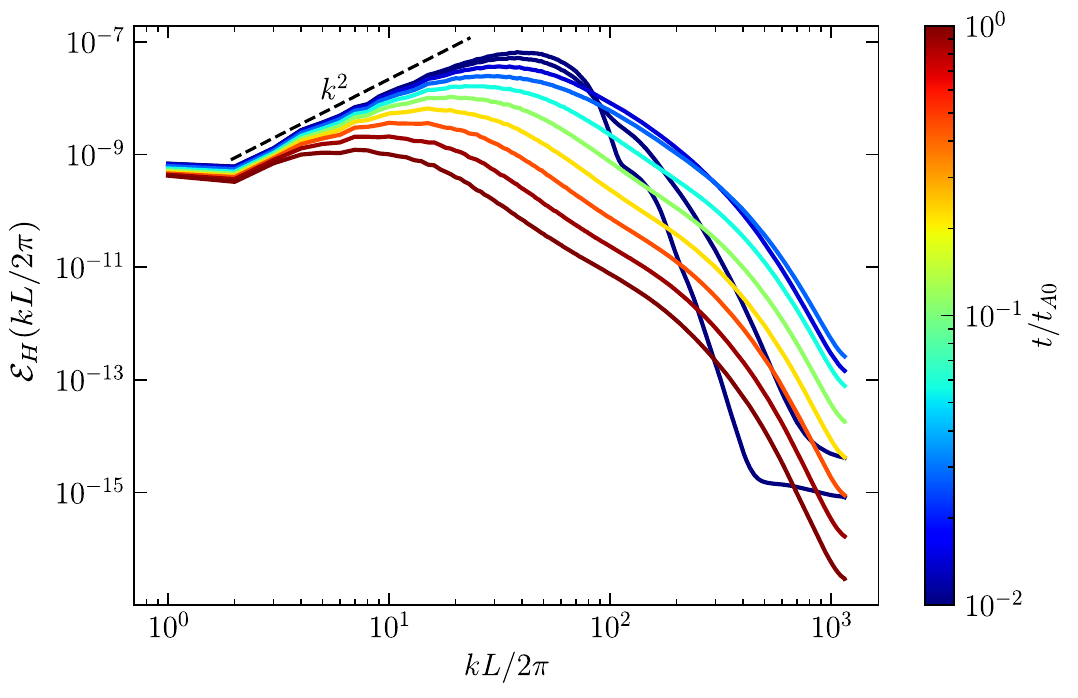}
\caption{\label{fig:IH_spectra}}
\end{subfigure}
\caption{\label{fig:emag_ekin_spectra}
Temporal and spectral statistics of decaying MHD turbulence in the simulations with $2304^3$ grid cells. Panel (a): magnetic and kinetic energy, $E_M$ and $E_K$, as functions of time in units of the Alfv\'en box-crossing time $t_{A0}$. The dotted vertical lines indicate times in which the correlation functions are plotted in figures~\ref{fig:unscaled_correlators} and \ref{fig:scaled_correlators}. Panel (b): magnetic- and kinetic-energy spectra. The darkest lines correspond to the initial condition, and the other three to the times indicated in panel~(a), with lighter lines indicating later times. We observe a $k^4$ scaling until the peak scale, after which the spectrum is proportional to $k^{-2}$, which may be an artifact of current-sheet discontinuities. Panel (c): magnetic-helicity-variance spectrum, $\mathcal{E}_H$. \blue{The vertical axis of each panel uses the code units defined in Section~\ref{sec:initconds}.} }
\end{figure}

\begin{figure}
    \centering
    \includegraphics[width=0.8\linewidth]{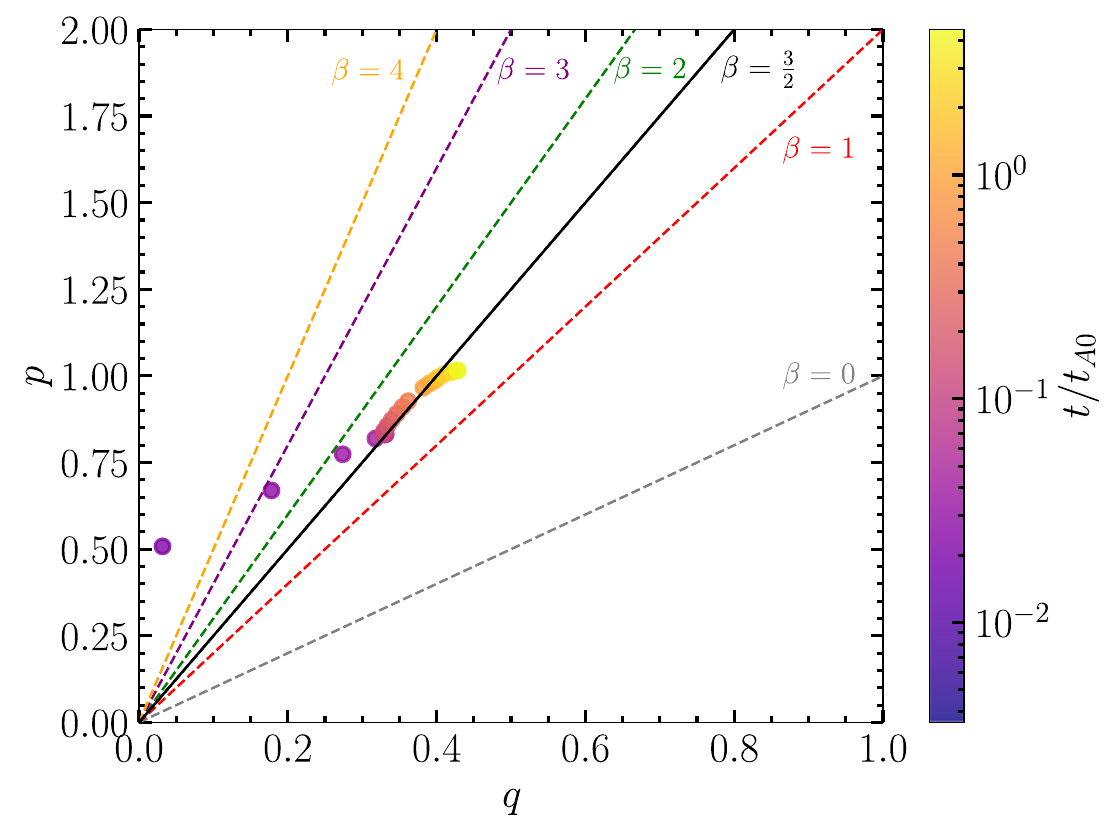}
    \caption{The  evolution of $p(t)$ and $q(t)$~\eqref{pq} as a function of time in the simulation. Different values of~$\beta$ correspond to different scaling relations between $E_{M}$ and $\xi_M$. The simulation evolves somewhat along the line $\beta = 3/2$, which corresponds to self-similar decay that conserves $I_H$ [see~\eqref{nonhelicallaws}].}
    \label{fig:pq_diagram}
\end{figure}

Figure~\ref{fig:emag_ekin_spectra} shows the  \blue{evolution of the magnetic and kinetic energies and their spectra, as well as that of the variance of magnetic helicity}. Both energies decay as a power law that is close to $E_{M}\propto t^{-1}$ (figure~\ref{fig:ekinemag}). \blue{In order to diagnose the decay laws more precisely, we }follow~\cite{Brandenburg17} in plotting in figure~\ref{fig:pq_diagram} the instantaneous scaling exponents of the magnetic energy and correlation length,
\begin{equation}
    p = -\frac{d \ln{E_{M}}}{d \ln{t}}, \quad q = \frac{d \ln{\xi_M}}{d\ln{t}}.\label{pq}
\end{equation}Both $p$ and $q$ increase with time somewhat, reaching $p\simeq1.02$ and $q \simeq 0.43$ at the last time shown. These values are close to the expected $p = 10/9 \simeq 1.11$ and $q = 4/9 \simeq 0.44$~\eqref{nonhelicallaws}. In particular, their evolution always satisfies $\beta \equiv p/q - 1 \simeq 3/2$, as is consistent with self-similar decay that conserves $I_H \sim E_{M}^2\xi_M^5$.\footnote{Consideration of the quantity $\beta$ is motivated by the fact that, for self-similar decay,
$\mathcal{E}_M(k, t)=\xi_M^{-\beta} \phi(k \xi_M)$, for some function $\phi(x)$ and constant $\beta$ \citep{Olesen97}.} 

Figure~\ref{fig:ekinemag} shows that the velocity field remains energetically subdominant to the magnetic field throughout the evolution\blue{, which may be a consequence of the velocity field being more intermittent than the volume-filling magnetic field, being concentrated in Alfv\'{e}nic reconnection outflows (see \citealt{HoskingSchekochihin20decay} for further discussion)}. Figure~\ref{fig:spectra} shows the evolution of the energy spectra. The magnetic spectrum \blue{$\mathcal{E}_M(k)$}  exhibits a near-$k^{-2}$ power law between its peak and the dissipation scale, which may be a signature of current-sheet discontinuities. The kinetic-energy spectrum $\mathcal{E}_K(k)$ is nearly flat over the same range, which may be associated with the sheet-like structure of the reconnection outflows \citep{HoskingSchekochihin20decay}. At large scales, the magnetic spectrum exhibits a $k^4$ tail that grows in amplitude with time [this is the ``inverse transfer'' effect discovered by~\cite{Brandenburg15} and \cite{Zrake14}, which was explained as a kinematic consequence of the conservation of $I_H$ by \cite{HoskingSchekochihin20decay}]; the \blue{amplitude of the} $k^4$ tail of the kinetic spectrum also grows somewhat over time. We show in figure~\ref{fig:IH_spectra} the spectrum of the variance of the magnetic-helicity density at different times. The conservation of its $k^2$ tail indicates the conservation of $I_H$~\citep{HoskingSchekochihin20decay}. 

\begin{figure}
    \centering
    \includegraphics[width=\linewidth]{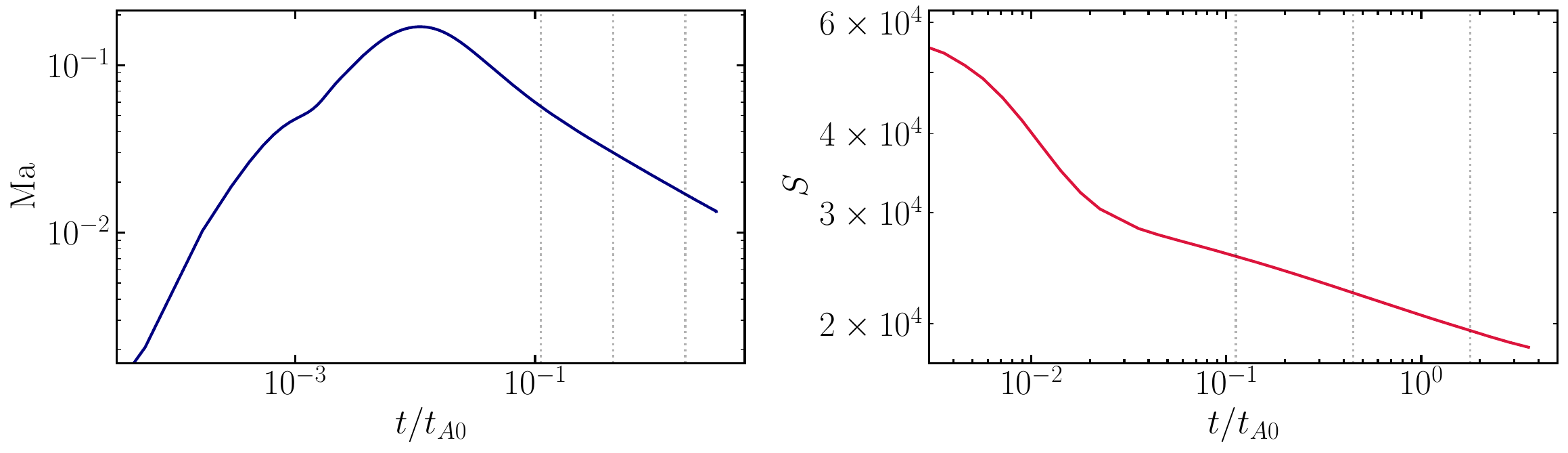}
    \caption{Evolution of the Mach number $\mach$ and Lundquist number $\lundquist$. The vertical lines indicate the times at which the correlation functions are shown in figures~\ref{fig:unscaled_correlators} and~\ref{fig:scaled_correlators}.}
    \label{fig:lundquist_mach}
\end{figure}

As we show in figure~\ref{fig:lundquist_mach}, the Lundquist number $\lundquist = v_A \xi_M/\eta$ decreases slowly with time [also as predicted by the decay laws~\eqref{nonhelicallaws}] but remains $\sim 10^{4}$. Our simulations are therefore marginal with respect to plasmoid-mediated fast reconnection \blue{of  magnetic structures at the integral scale} (see~\citealt{Uzdensky10}\footnote{\blue{We note that recent numerical results have indicated that very high resolution and robust seeding are needed to properly resolve plasmoid-mediated fast reconnection in undriven 2D MHD---see~\citealt{MorilloAlexakis25} and~\citealt{Vicentin25}.}}). \blue{The velocity Reynolds number (not shown) based on the integral scale of the velocity field [i.e., \eqref{xi_M}, but evaluated for the velocity spectrum] peaks at $\simeq 2\times 10^4$ at $t/t_{A0}\simeq 10^{-2}$, and then  decays by a factor of $\simeq 3$ by $t/t_{A0}\simeq 3$.} Figure~\ref{fig:lundquist_mach} shows that the Mach number $\mach \equiv \langle v^2\rangle^{1/2}/c_s \lesssim 0.1$ at all times, making the turbulence nearly incompressible throughout.

\subsection{Measurements of correlation functions\label{sec:measurements}}

\begin{figure}
    \centering
    \includegraphics[width = 0.7\linewidth]{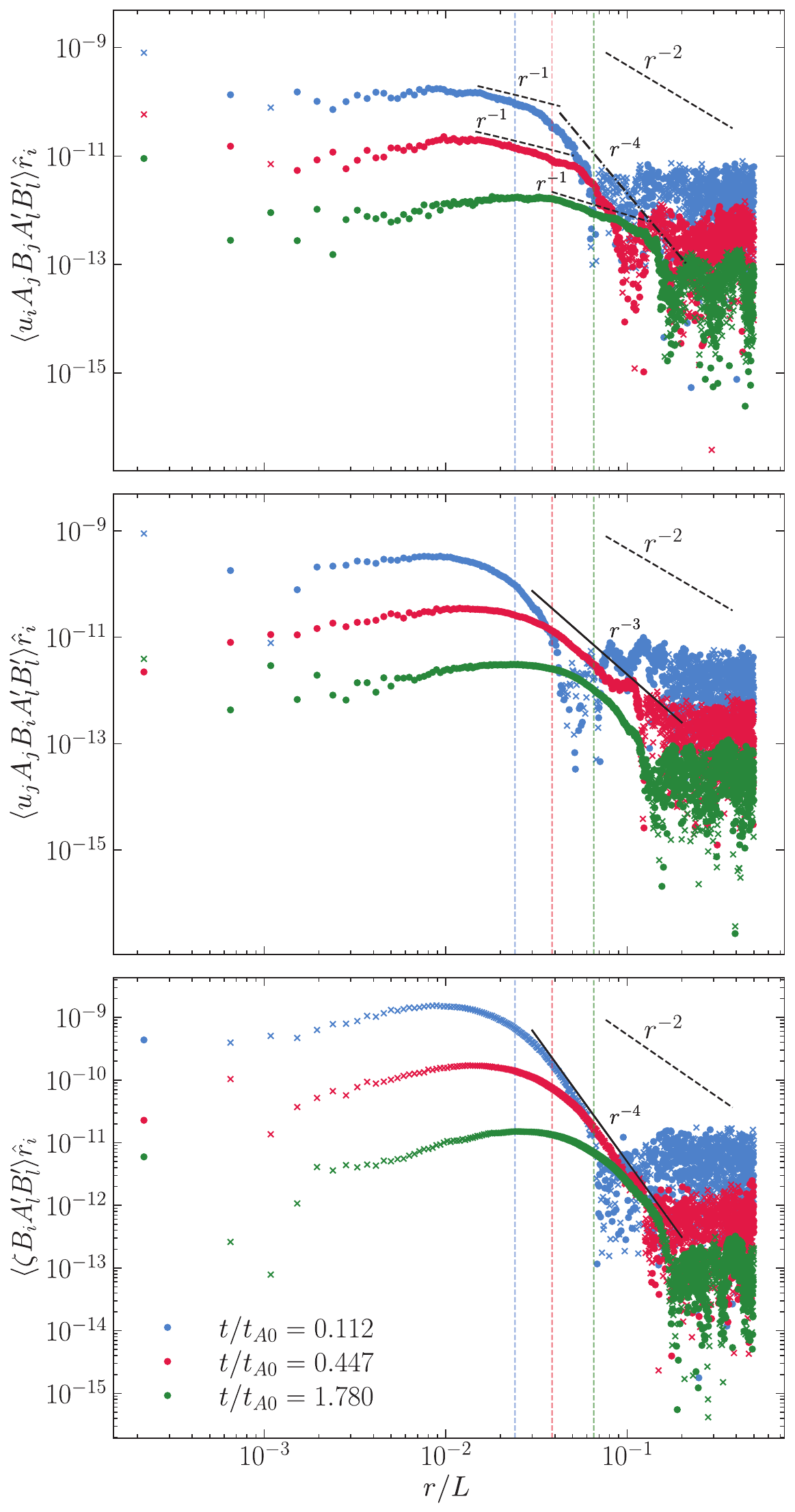}
    \caption{The three correlation functions that contribute to $C_{\infty}$~\eqref{Cinfty} measured at ${t/t_{A0} = 0.112}$ (blue), $t/t_{A0} = 0.447$ (red) and $t/t_{A0} = 1.780$ (green) as a function of $r/L$. \blue{Each vertical axis is measured in the code units defined in Section~\ref{sec:initconds}.} Positive values are plotted with solid circles, negative ones with crosses. Dashed vertical lines indicate the integral scale $\xi_M$~\eqref{xi_M} at each time. The correlation functions decay in amplitude and shift to larger spatial scales over time. We indicate a number of power laws to guide the eye, but observe that none of these fit any given curve over more than half a decade in $r/L$.}
    \label{fig:unscaled_correlators}
\end{figure}

We now describe our measurements of the correlation functions that appear in~\eqref{Cinfty}. We calculate these using an analogous method to the calculation of higher-order structure functions by \cite{federrath2021sonic}, using $10^{11}$ sampling points. We plot the correlation functions in figure~\ref{fig:unscaled_correlators} for $t/t_{A0}$ of $0.112$, $0.447$ and $1.780$, at which times the magnetic energy has decayed by factors of roughly $10$, $30$ and $100$, respectively (see figure~\ref{fig:ekinemag}). \blue{In each case, the correlation functions vanish as $r \to 0$ (where they become dominated by noise). This is as expected, since the expectation value of a vector is zero in isotropic turbulence. The correlation functions peak in magnitude at around $r\sim \xi_M$}. At ${r/\xi_M \gtrsim 1}$, their amplitudes decay with $r$ in what we term the ``decorrelation range'', ultimately becoming dominated by numerical noise. We observe that they decrease in amplitude and shift to larger $r$ over time, consistent with the decay of the turbulence and transfer of energy to larger scales [see the decay laws~\eqref{nonhelicallaws} and figure~\ref{fig:emag_ekin_spectra}].

In the decorrelation range $r/\xi_M \gtrsim 1$, each correlation function ultimately decays faster than $r^{-2}$, which is the condition for the conservation of $I_H$ [$C_{\infty}=0$ in~\eqref{dIHdt}] (we note that the top panel of figure~\ref{fig:unscaled_correlators} appears to show $r^{-1}$ over a short intermediate range, but this steepens at larger $r$). It is difficult to judge whether any of these plots show a power-law decay at $r \gg \xi_M$: the $r^{-3}$ and $r^{-4}$ power laws that we plot for reference match our measurements reasonably well locally, although this is true only for around half a decade in $r$. After this range, the behaviour transitions to steeper decay in most cases. Because the Coulomb gauge is of the Poisson type [see \eqref{poissongauge} and~\eqref{coulomb_zeta}], the analysis in Section~\ref{sec:calculation} \blue{rules out decay of these correlation functions slower than $r^{-6}$ [which corresponds to terms in the Taylor expansion that involve both pressure and gauge, i.e., counting two inverse powers of $r$ from~\eqref{coulomb_zeta} and four from the gradient of~\eqref{eqn:pressure_theoretical}]}. Supposing that the $r^{-3}$ and $r^{-4}$ power laws identified in figure~\ref{fig:unscaled_correlators} are real, therefore, they are somewhat less steep than our theoretical analysis predicts. One way to rationalise this behaviour is by noting that our analysis involved frequent use of the vanishing of terms like $\langle \vb \bcdot \bb \rangle$ and $\langle \ab \bcdot \bb \rangle$ due to parity symmetry. While these quantities would be zero in an infinite domain, the box-averaged $\vb\bcdot \bb$ and $\ab\bcdot \bb$ in a finite periodic box can be non-zero, owing to fluctuations sourced by compressibility and resistivity. Thus, parity-dependent terms that would, if non-zero, contribute shallower power laws to the various correlation functions [such as~\eqref{integral1}, \eqref{integral2} and \eqref{integral3}], might, at $r\gg \xi_M$, overwhelm the faster-decaying parity-invariant contributions.

\begin{figure}
    \centering
    \includegraphics[width = \textwidth]{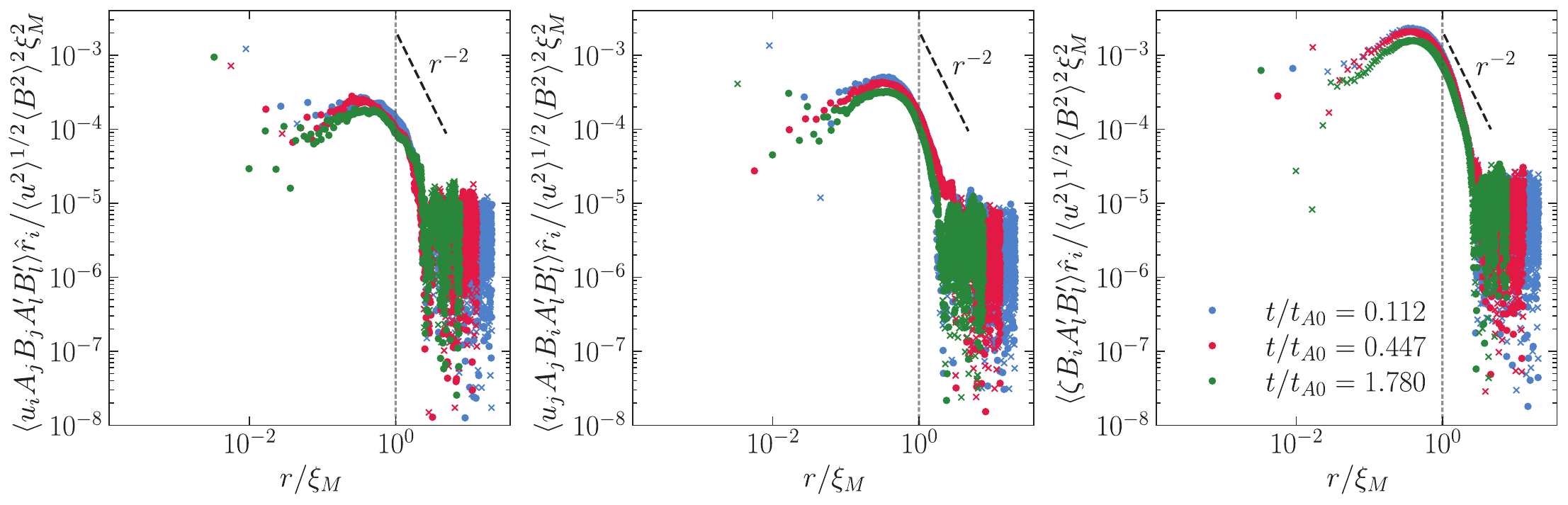}
    \caption{The same as figure~\ref{fig:unscaled_correlators}, but with all quantities normalised to the dimensionally appropriate combinations of $\langle u^2\rangle^{1/2}$, $\langle B^2\rangle^{1/2}$ and $\xi_M$.
    \label{fig:scaled_correlators}}
\end{figure}

In figure~\ref{fig:scaled_correlators}, we show the same correlation functions as in figure~\ref{fig:unscaled_correlators}, but normalised to the dimensionally appropriate combinations of $\langle u^2\rangle^{1/2}$, $\langle B^2\rangle^{1/2}$ and $\xi_M$. We find, under this rescaling, the functions collapse onto one another and become nearly time-independent, as is consistent with self-similar decay.

\section{Conclusion\label{sec:conclusion}}

In this work, we have examined the theoretical justification for the conservation of the integral $I_H$~\eqref{I_H} in non-helical, isotropic MHD turbulence in considerably greater detail than in previous studies. We have employed the methodology of \cite{BatchelorProudman56} to determine whether the boundary term~$C_{\infty}$ in the evolution equation for $I_H$~\eqref{dIHdt} vanishes for decay from an initial condition for which distant points are statistically independent. We find that $I_H$ is conserved ($C_{\infty} = 0$) for a wide class of local and non-local gauge functions appearing in the evolution equation for the magnetic vector potential~\eqref{dAdt}. For the case of the Coulomb gauge $\bnabla \bcdot \ab=0$, we have measured the relevant correlation functions directly in a simulation of decaying turbulence (Section~\ref{sec:measurements}), and found that, indeed, $C_{\infty}=0$ (although the correlation functions appear to decay somewhat more slowly than predicted by our theory, possibly due to box-scale fluctuations in parity-dependent quantities). 

While our theory predicts that $I_H$ is conserved under the wide class of gauge choices described above, we have nonetheless also identified a simple (but, to the best of our knowledge, previously unconsidered) class of gauges~\eqref{funkygauge} for which our theory indicates that $C_{\infty}$ can be finite, and, therefore, $I_H$ is not conserved. It is our conjecture that, in transformation to such a gauge, magnetic-helicity density is redistributed along wandering field lines in such a way that its fluctuation level no longer encodes the relevant topological constraints on a local patch of tangled magnetic field. This would be the case if helicity density were exchanged between points that were not causally connected, for example. We plan to explore this aspect further in future work.

\section{Acknowledgements}
We are grateful to H. Politano for helpful advice. We also thank S.~Banerjee, A.~Bhattacharjee, R.~W.~Boswell, A.~Brandenburg, R.~L.~Dewar, S.~Galtier, Z.~Hemler, P.~A.~Kempski, M.~W.~Kunz, H.~N.~Latter, J.~Squire and R.~Wielian for helpful discussions. D.~N.~H. is grateful for many conversations about decaying turbulence with A.~A.~Schekochihin. \blue{We are also grateful to two anonymous reviewers, whose comments greatly improved this manuscript.} J.~K.~J.~H.~acknowledges funding via the Bok Honours Scholarship, the Space Plasma, Astronomy and Astrophysics Research Award and the Boswell Technologies Endowment Fund. C.~F.~acknowledges funding provided by the Australian Research Council (Discovery Projects DP230102280 and DP250101526), and the Australia-Germany Joint Research Cooperation Scheme (UA-DAAD). J.~R.~B.~acknowledges financial support from the Australian National University, via the Deakin PhD and Dean's Higher Degree Research (theoretical physics) Scholarships, the Australian Government via the Australian Government Research Training Program Fee-Offset Scholarship and the Australian Capital Territory Government funded Fulbright scholarship. We~further acknowledge high-performance computing resources provided by the Leibniz Rechenzentrum and the Gauss Centre for Supercomputing (grants~pr32lo, pr48pi and GCS Large-scale project~10391), the Australian National Computational Infrastructure in the framework of the National Computational Merit Allocation Scheme, ANU Merit Allocation Scheme (grant~ek9), and the ANU Startup Scheme (grant~xx52), as well as the Pawsey Supercomputing Centre (grant~pawsey0810).

\appendix

\color{black}\section{Derivation of equation~\eqref{I_Hgaugetransformation}}\label{app:gauge-IH}

Let $\chi(r) \equiv \langle h(\xb)h(\xb+\rb)\rangle \equiv \langle hh'\rangle$, where primes denote fields evaluated at $\xb'= \xb + \rb$,  be the helicity-density correlation function that appears inside $I_H$~\eqref{I_H}, i.e.,
\begin{equation}
I_H = \int_{\blue{\mathbb{R}^3}} d^3\rb \chi(r).
\end{equation}
Under a gauge transformation $\Ab\to \Ab+\bnabla\varphi$,
\begin{equation}
h\to h+\Bb\bcdot \bnabla\varphi = h+\bnabla\bcdot (\varphi\Bb ),
\end{equation}
and similarly at $\xb'=\xb+\rb$. Hence, $\chi \to \chi + \delta \chi$, where
\begin{align}
\delta \chi
&= \langle [\bnabla\bcdot(\varphi\Bb)]h'\rangle
  + \langle h[\bnabla'\bcdot(\varphi'\Bb')]\rangle
  + \langle [\bnabla\bcdot(\varphi\Bb)][\bnabla'\bcdot(\varphi'\Bb')]\rangle.\label{eq:deltaC-start}
\end{align}
Using linearity of averaging and homogeneity,
\begin{equation}
\partial_{x_i}\langle \cdots \rangle = -\partial_{r_i}\langle \cdots \rangle,
\qquad
\partial_{x'_j}\langle \cdots \rangle = \phantom{-}\partial_{r_j}\langle \cdots \rangle,
\end{equation}
and commuting derivatives with the average, \eqref{eq:deltaC-start} becomes
\begin{equation}
\delta \chi
= -\partial_{r_i}\langle \varphi B_ih'\rangle
   + \partial_{r_j}\langle h\varphi'B_j'\rangle
   - \partial_{r_i}\partial_{r_j}\langle \varphi B_i\varphi'B_j'\rangle.\label{eq:deltachi}
\end{equation}By homogeneity and isotropy, the vectors 
{\(
\mathcal{F}_i(\rb)\equiv\langle h\varphi' B'_i\rangle
\)}
and ${\mathcal{G}_i(\rb)\equiv\langle \varphi B_ih'\rangle}$ must have the form $\hat r_i F(r)$ and $\hat r_i G(r)$, respectively, for some functions $F$ and $G$. Further, homogeneity implies
$\mathcal{G}_i(\rb)=\mathcal{F}_i(-\rb)$, whence ${G(r)=-F(r)}$.
Therefore,
\begin{equation}
 \langle h\varphi' B'_i\rangle
= -  \langle \varphi B_ih'\rangle.\label{appA_symmetry}
\end{equation}
Integrating~\eqref{eq:deltachi} over $\rb$ and using~\eqref{appA_symmetry}, we find that $I_H\to I_H+\delta I_H$ under the gauge transformation, where
\begin{equation}
\delta I_H = \int_{\blue{\mathbb{R}^3}} d^3\rb\delta \chi(\rb) = -2\int_{\blue{\mathbb{R}^3}} d^3\rb\partial_{r_i}\langle \varphi B_ih'\rangle
   - \int_{\blue{\mathbb{R}^3}} d^3\rb\partial_{r_i}\partial_{r_j}\langle \varphi B_i\varphi' B_j'\rangle.\label{eq:deltaI-divform}
\end{equation}

The first integral in \eqref{eq:deltaI-divform} is
\begin{equation}
\lim_{R\to\infty}\int_{B_R} \partial_{r_i}\langle \varphi B_ih'\rangle d^3\rb = \lim_{R\to\infty}\int_{S_R} \langle \varphi B_ih'\rangle\hat r_i dS
 = \lim_{R\to\infty}4\pi R^2 \theta(R),\label{eq:appA_int1}
\end{equation}where $B_R$ is a sphere of radius $R$, $S_R$ is its boundary, and $\theta(r)\equiv \langle \varphi B_i h'\rangle \hat{r}_i$. For the second integral in~\eqref{eq:deltaI-divform}, we introduce the standard isotropic decomposition
\begin{equation}
T_{ij}(\rb)\equiv\langle \varphi B_i\varphi' B_j'\rangle
= f(r)\delta_{ij}+g(r)\hat r_i \hat r_j,\label{eq:iso-T}
\end{equation}where $\hat{r}_i\equiv r_i/r$.
Using $\partial_{r_j}\hat r_i=(\delta_{ij}-\hat r_i\hat r_j)/r$ and $\partial_{r_j}\hat r_j=2/r$, we have
\begin{equation}
\hat r_i\partial_{r_j}T_{ij} = f'(r)+g'(r)+\frac{2g(r)}{r}.
\end{equation}The second integral in~\eqref{eq:deltaI-divform} may therefore be expressed as
\begin{align}
\lim_{R\to\infty}\int_{B_R}\partial_{r_i}\partial_{r_j}T_{ij}d^3\rb
&= \lim_{R\to\infty}\int_{S_R} \hat r_i\partial_{r_j}T_{ij}dS
= \lim_{R\to\infty}4\pi R^2\left[f'(R)+g'(R)+\frac{2g(R)}{R}\right]\nonumber\\
&= \lim_{R\to\infty}4\pi\left[ R^2 f'(R) + \frac{d}{dR}\big(R^2 g(R)\big)\right].\label{eq:double-div}
\end{align}Substituting~\eqref{eq:appA_int1} and~\eqref{eq:double-div} into~\eqref{eq:deltaI-divform} yields
\begin{equation}
\delta I_H
= -4\pi\lim_{r\to\infty}\left[
2r^2\theta(r)+r^2\,f'(r)+\frac{d}{dr}\big(r^2 g(r)\big)
\right],\label{eq:IH-gauge-final}
\end{equation}which is the same as~\eqref{I_Hgaugetransformation}.

\section{Definition of cumulants\label{app:cumulants}}

The $n$th-order cumulant (Ursell function) of the random variables $X_1, \dots, X_n$ (in this work, these might represent a given component of the velocity or magnetic field at a particular point in space, for example) is given by (see, e.g.,~\citealt{McCullagh18})
\begin{equation}
    \langle X_1 \cdots X_n \rangle_c
= \left.\frac{\partial^n}{\partial z_1 \cdots \partial z_n}
\ln \left\langle \exp\left(\sum_{j=1}^n z_j X_j\right) \right\rangle
\right|_{z_1=\cdots=z_n=0}.
\end{equation}In the case of $n=4$, for example,
\begin{align}
\langle X_1 X_2 X_3 X_4\rangle_c
&= \langle X_1 X_2 X_3 X_4\rangle
- \langle X_1 X_2\rangle \langle X_3 X_4\rangle
- \langle X_1 X_3\rangle \langle X_2 X_4\rangle
- \langle X_1 X_4\rangle \langle X_2 X_3\rangle\nonumber\\
&\quad
- \langle X_1 X_2 X_3\rangle \langle X_4\rangle
- \langle X_1 X_2 X_4\rangle \langle X_3\rangle
- \langle X_1 X_3 X_4\rangle \langle X_2\rangle
- \langle X_2 X_3 X_4\rangle \langle X_1\rangle\nonumber\\
&\quad
+ 2\big(
\langle X_1 X_2\rangle \langle X_3\rangle \langle X_4\rangle
+ \langle X_1 X_3\rangle \langle X_2\rangle \langle X_4\rangle
+ \langle X_1 X_4\rangle \langle X_2\rangle \langle X_3\rangle\nonumber\\
&\qquad\ 
+ \langle X_2 X_3\rangle \langle X_1\rangle \langle X_4\rangle
+ \langle X_2 X_4\rangle \langle X_1\rangle \langle X_3\rangle
+ \langle X_3 X_4\rangle \langle X_1\rangle \langle X_2\rangle
\big)\nonumber\\
&\quad
- 6\,\langle X_1\rangle \langle X_2\rangle \langle X_3\rangle \langle X_4\rangle,
\end{align}which reduces to
\begin{equation}
    \langle X_1 X_2 X_3 X_4\rangle_c
= \langle X_1 X_2 X_3 X_4\rangle
- \langle X_1 X_2\rangle \langle X_3 X_4\rangle
- \langle X_1 X_3\rangle \langle X_2 X_4\rangle
- \langle X_1 X_4\rangle \langle X_2 X_3\rangle
\end{equation}if the random variables $X_i$ have zero mean. The cumulant vanishes if the random variables $X_1, \dots, X_n$ can be divided into two non-empty independent sets.

\section{The case of a vector potential with mixed parity}

In Section~\ref{sec:calculation}, we made frequent use of the fact that the vector potential $\ab$, being related to the axial (pseudo-)vector $\bb$ by $\bb = \bnabla \times \ab$, is a polar (true) vector. In principle, this may not be true---one is always free to add to $\ab$ the gradient of a quantity that is not a true-scalar, in which case, the new $\ab$ will have mixed parity. For completeness, we here point out how the conclusions of Section~\ref{sec:calculation} are modified if this is the case.

If the correlation function~$\langle B_r^{\prime\prime} B_s^{\prime\prime} A^{\prime}_l B^\prime_l\rangle$ does not vanish by parity symmetry,~\eqref{dt3_5thcorrelator} becomes, after use of the large-$r$ expansion~\eqref{s_expansion},
\begin{multline}
    \frac{\p^2 }{\p t^2}\langle u_iA_jB_kA_l'B_l'\rangle = 
     \dots +\bigg\langle  \frac{\p u_i}{\p x_m} A_j B_k \bigg\rangle \frac{\p^3}{\p r_m \p r_s \p r_r} \left(\frac{1}{r}\right)\\\times\frac{1}{4\pi} \int_{\blue{\mathbb{R}^3}} d^3 \Sb^\prime\,\langle B_r^{\prime\prime} B_s^{\prime\prime} A^{\prime}_l B^\prime_l\rangle + \mathcal{O}\left(\frac{1}{r^5}\right) +\dots\label{final_dt2_5th}
\end{multline}The term that we have isolated in~\eqref{final_dt2_5th} vanishes if $\ab$ is a polar (true) vector, but otherwise appears to be $\mathcal{O}(r^{-4})$. However, the mixed-parity $\ab$ can always be related to a polar vector potential by a gauge transformation, so $A_l'$ can be replaced by $\p \varphi' /\p x_l' $ inside the correlation function. The derivative can be brought outside the correlation function, making the particular term under consideration $\mathcal{O}(r^{-5})$. This vanishing of the $\mathcal{O}(r^{-4})$ term relied on the contraction of $A_l^\prime$ with a divergence-free field in~\eqref{final_dt2_5th}---this will not be the case for all terms contributing to the Taylor-series expansion of $\langle u_iA_jB_kA_l'B_l'\rangle$, so in general, we expect $\langle u_iA_jB_kA_l'B_l'\rangle=\mathcal{O}(r^{-4})$. Analogous reasoning implies that $\langle \zeta B_iA_l'B_l'\rangle = \mathcal{O}(r^{-4})$ if $\ab$ is not a true vector.

We conclude that if the vector potential has mixed parity, symmetry properties do not guarantee as rapid a decay of the helicity-flux correlation function, but the conclusion that $C_{\infty} = 0$ is not violated.
\color{black}

\section{Resolution study} \label{sec:resolution_study}
\begin{figure}
    \centering
    \includegraphics[width = \linewidth]{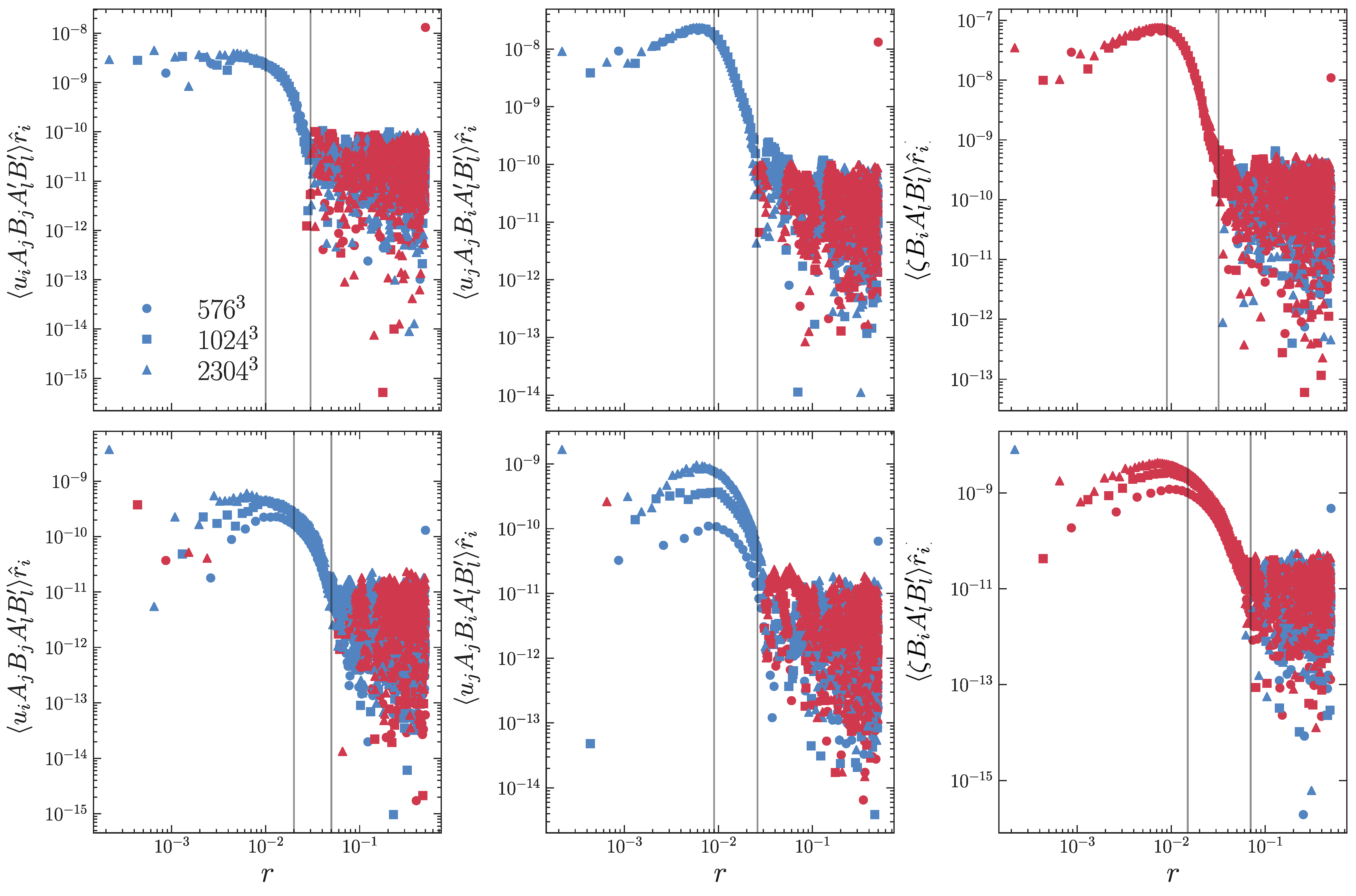}
    \caption{We plot the three different correlators that appear in~\eqref{Cinfty} at three different resolutions, at $t/t_{A0} = 0.018$ (upper panels) and $t/t_{A0} = {0.056}$ (lower panels). Blue points correspond to data with positive values, red points to the absolute values of data with negative values.}
    \label{fig:convergence_study}
\end{figure}

We have checked the resolution dependence of the results reported in Section~\ref{sec:numerics} by comparing them with analogous measurements from simulations with resolutions of $576^3$ and $1024^3$. We show in figure~\ref{fig:convergence_study} the correlation functions shown in figures~\ref{fig:unscaled_correlators} and~\ref{fig:scaled_correlators} at $t/t_{A0} = 0.018$ and $t/t_{A0} = 0.056$. The correlation functions are essentially identical between simulations at $t/t_{A0} = 0.018$ (this is, essentially, the initial condition), but some deviation exists by $t/t_{A0} = 0.056$. Nonetheless, the correlation functions appear converged within the decorrelation range, indicating that our results in Section~\ref{sec:measurements} are converged at $2304^3$. 

\section{Decay at smaller Lundquist numbers\label{lowerLundquist}}
\begin{figure}
    \centering
    \includegraphics[width = 0.5\textwidth]{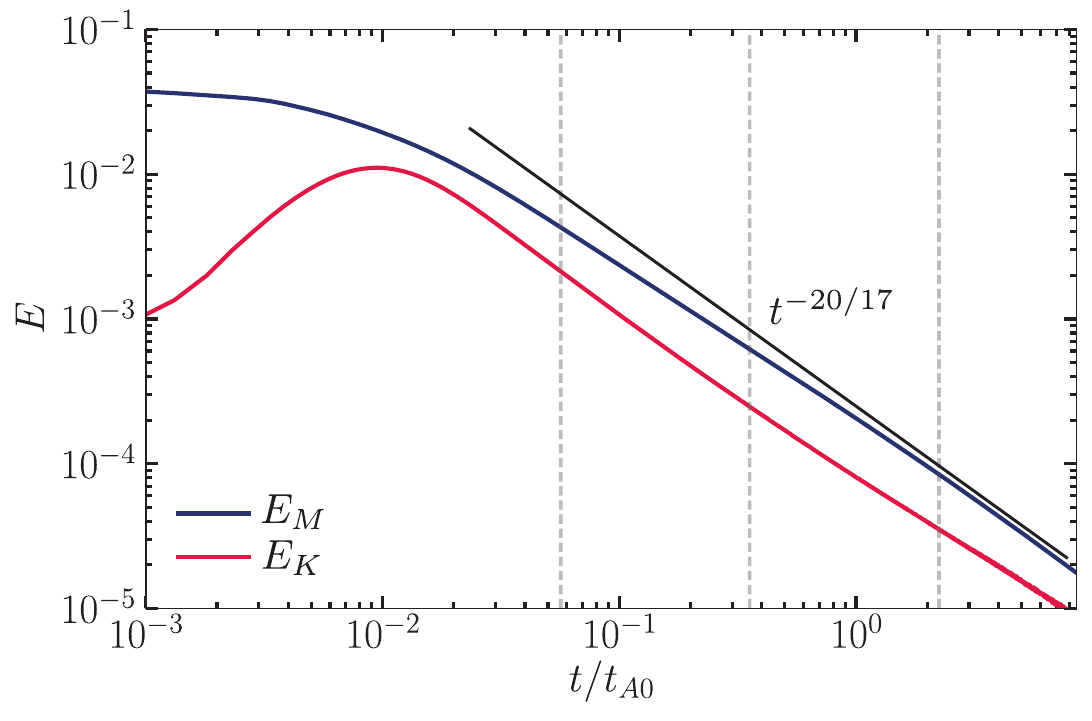}
    \caption{Temporal evolution of the magnetic and kinetic energies for a $1204^3$ simulation. }
    \label{fig:emag_ekin_1204}
\end{figure}

We show in figure~\ref{fig:emag_ekin_1204} the evolution of the magnetic and kinetic energies in a $1204^3$ simulation with initial Lundquist number $\lundquist_0 \sim 10^3$, well below the critical value of $10^4$ for plasmoid-mediated fast reconnection. We find a power-law decay of these quantities, with \blue{$E_{M}$  close to  $t^{-20/17}$ over 1-2 decades} in $t/t_{A0}$. This \blue{power law is} as expected for self-similar decay that conserves $I_H$ and happens on the Sweet-Parker-reconnection timescale~\citep{HoskingSchekochihin20decay}. Deviation from this decay law at higher Lundquist number (see figure~\ref{fig:ekinemag}) may be a consequence of transition to an $\eta$-independent reconnection timescale, although other possibilities have been mooted---see~\citealt{Zhou22_Hosking} and~\citealt{Brandenburg24b}.

\bibliographystyle{jpp}

\newcommand{\apj}{The Astrophysical Journal}
\newcommand{\apjs}{{\apj} Suppl.~Series}
\newcommand{\aap}{Astronomy \& Astrophysics}
\newcommand{\jcp}{Journal of Computational Physics}
\bibliography{jpp-instructions,federrath, decay_mod}

\begin{thebibliography}{51}
\expandafter\ifx\csname natexlab\endcsname\relax\def\natexlab#1{#1}\fi
\def\au#1{#1} \def\ed#1{#1} \def\yr#1{#1}\def\at#1{#1}\def\jt#1{\textit{#1}}
  \def\bt#1{#1}\def\bvol#1{\textbf{#1}} \def\vol#1{#1} \def\pg#1{#1}
  \def\publ#1{#1}\def\arxiv#1{#1}\def\org#1{#1}\def\st#1{\textit{#1}}

\bibitem[{Banerjee} \& {Jedamzik}(2004)]{BanerjeeJedamzik04}
{\sc \au{{Banerjee}, R.} \& \au{{Jedamzik}, K.}} \yr{2004}  \at{{Evolution of
  cosmic magnetic fields: From the very early Universe, to recombination, to
  the present}}.  \jt{Phys. Rev. D}  \bvol{70},  \pg{123003}.

\bibitem[Batchelor(1953)]{batchelor1953THT}
{\sc \au{Batchelor, {G.~K.}}} \yr{1953} {\em The theory of homogeneous
  turbulence\/}.  \publ{Cambridge: Cambridge University Press}.

\bibitem[{Batchelor} \& {Proudman}(1956)]{BatchelorProudman56}
{\sc \au{{Batchelor}, G.~K.} \& \au{{Proudman}, I.}} \yr{1956}  \at{{The
  large-scale structure of homogeneous turbulence}}.  \jt{Philos. Trans. R.
  Soc. A}  \bvol{248},  \pg{369}.

\bibitem[{Berger}(1984)]{Berger84}
{\sc \au{{Berger}, M.~A.}} \yr{1984}  \at{{Rigorous new limits on magnetic
  helicity dissipation in the solar corona}}.  \jt{Geophys. Astrophys. Fluid
  Dyn.}  \bvol{30},  \pg{79}.

\bibitem[{Berger}(1997)]{Berger97}
{\sc \au{{Berger}, M.~A.}} \yr{1997}  \at{{Magnetic helicity in a periodic
  domain}}.  \jt{J. Geophys. Res.}  \bvol{102},  \pg{2637}.

\bibitem[{Bhat} {\em et~al.\/}(2021){Bhat}, {Zhou} \& {Loureiro}]{Bhat21}
{\sc \au{{Bhat}, P.}, \au{{Zhou}, M.} \& \au{{Loureiro}, N.~F.}} \yr{2021}
  \at{{Inverse energy transfer in decaying, three-dimensional, non-helical
  magnetic turbulence due to magnetic reconnection}}.  \jt{Mon. Not. R. Astron.
  Soc.}  \bvol{501},  \pg{3074}.

\bibitem[{Biskamp} \& {M{\"u}ller}(1999)]{BiskampMuller99}
{\sc \au{{Biskamp}, D.} \& \au{{M{\"u}ller}, W.-C.}} \yr{1999}  \at{{Decay laws
  for three-dimensional magnetohydrodynamic turbulence}}.  \jt{Phys. Rev.
  Lett.}  \bvol{83},  \pg{2195}.

\bibitem[Bouchut {\em et~al.\/}(2007)Bouchut, Klingenberg \&
  Waagan]{bouchut2007multiwave}
{\sc \au{Bouchut, F.}, \au{Klingenberg, C.} \& \au{Waagan, K.}} \yr{2007}
  \at{A multiwave approximate riemann solver for ideal mhd based on relaxation.
  i: theoretical framework}.  \jt{Numerische Mathematik}  \bvol{108},
  \pg{7--42}.

\bibitem[Bouchut {\em et~al.\/}(2010)Bouchut, Klingenberg \&
  Waagan]{bouchut2010multiwave}
{\sc \au{Bouchut, F.}, \au{Klingenberg, C.} \& \au{Waagan, K.}} \yr{2010}
  \at{A multiwave approximate riemann solver for ideal mhd based on relaxation
  ii: numerical implementation with 3 and 5 waves}.  \jt{Numerische Mathematik}
   \bvol{115},  \pg{647--679}.

\bibitem[{Brandenburg}(2023)]{Brandenburg23c}
{\sc \au{{Brandenburg}, A.}} \yr{2023}  \at{{Hosking integral in non-helical
  Hall cascade}}.  \jt{J. Plasma Phys.}  \bvol{89},  \pg{175890101}.

\bibitem[{Brandenburg} \& {Banerjee}(2025)]{Brandenburg25_twoconserved}
{\sc \au{{Brandenburg}, A.} \& \au{{Banerjee}, A.}} \yr{2025}  \at{{Turbulent
  magnetic decay controlled by two conserved quantities}}.  \jt{J. Plasma
  Phys.}  \bvol{91},  \pg{E5}.

\bibitem[{Brandenburg} \& {Kahniashvili}(2017)]{Brandenburg17}
{\sc \au{{Brandenburg}, A.} \& \au{{Kahniashvili}, T.}} \yr{2017}  \at{{Classes
  of hydrodynamic and magnetohydrodynamic turbulent decay}}.  \jt{Phys. Rev.
  Lett.}  \bvol{118},  \pg{055102}.

\bibitem[{Brandenburg} {\em et~al.\/}(2015){Brandenburg}, {Kahniashvili} \&
  {Tevzadze}]{Brandenburg15}
{\sc \au{{Brandenburg}, A.}, \au{{Kahniashvili}, T.} \& \au{{Tevzadze}, A.~G.}}
  \yr{2015}  \at{{Nonhelical inverse transfer of a decaying turbulent magnetic
  field}}.  \jt{Phys. Rev. Lett.}  \bvol{114},  \pg{075001}.

\bibitem[{Brandenburg} {\em et~al.\/}(2023){Brandenburg}, {Kamada} \&
  {Schober}]{Brandenburg23a}
{\sc \au{{Brandenburg}, A.}, \au{{Kamada}, K.} \& \au{{Schober}, J.}} \yr{2023}
   \at{{Decay law of magnetic turbulence with helicity balanced by chiral
  fermions}}.  \jt{Phys. Rev. Res.}  \bvol{5},  \pg{L022028}.

\bibitem[{Brandenburg} \& {Larsson}(2023)]{Brandenburg23b}
{\sc \au{{Brandenburg}, A.} \& \au{{Larsson}, G.}} \yr{2023}  \at{{Turbulence
  with magnetic helicity that is absent on average}}.  \jt{Atmosphere}
  \bvol{14},  \pg{932}.

\bibitem[{Brandenburg} {\em et~al.\/}(2024){Brandenburg}, {Neronov} \&
  {Vazza}]{Brandenburg24b}
{\sc \au{{Brandenburg}, A.}, \au{{Neronov}, A.} \& \au{{Vazza}, F.}} \yr{2024}
  \at{{Resistively controlled primordial magnetic turbulence decay}}.
  \jt{Astron. Astrophys.}  \bvol{687},  \pg{A186}.

\bibitem[{Brandenburg} {\em et~al.\/}(2025){Brandenburg}, {Yi} \&
  {Wu}]{Brandenburg25_columnar}
{\sc \au{{Brandenburg}, A.}, \au{{Yi}, L.} \& \au{{Wu}, X.}} \yr{2025}
  \at{{Inverse cascade from helical and non-helical decaying columnar magnetic
  fields}}.  \jt{J. Plasma Phys.}  \bvol{91},  \pg{E113}.

\bibitem[Davidson(2000)]{davidson2000loitsyansky}
{\sc \au{Davidson, P.~A.}} \yr{2000}  \at{Was loitsyansky correct? a review of
  the arguments}.  \jt{Journal of Turbulence}  \bvol{1}~(1),  \pg{006}.

\bibitem[Davidson(2015)]{Davidson15}
{\sc \au{Davidson, P.~A.}} \yr{2015} {\em {Turbulence: an Introduction for
  Scientists and Engineers}\/}.  \publ{Oxford University Press}.

\bibitem[{Dubey} {\em et~al.\/}(2008){Dubey}, {Fisher}, {Graziani}, {Jordan},
  {Lamb}, {Reid}, {Rich}, {Sheeler}, {Townsley} \& {Weide}]{DubeyEtAl2008}
{\sc \au{{Dubey}, A.}, \au{{Fisher}, R.}, \au{{Graziani}, C.}, \au{{Jordan},
  IV, G.~C.}, \au{{Lamb}, D.~Q.}, \au{{Reid}, L.~B.}, \au{{Rich}, P.},
  \au{{Sheeler}, D.}, \au{{Townsley}, D.} \& \au{{Weide}, K.}} \yr{2008}
  {Challenges of Extreme Computing using the FLASH code}.  \bt{In {\em
  Numerical Modeling of Space Plasma Flows\/} (ed. \ed{N.~V. {Pogorelov},
  E.~{Audit} \& G.~P. {Zank}})},  \st{Astronomical Society of the Pacific
  Conference Series},  \vol{vol. 385},  \pg{p. 145}.

\bibitem[Federrath {\em et~al.\/}(2021)Federrath, Klessen, Iapichino \&
  Beattie]{federrath2021sonic}
{\sc \au{Federrath, C.}, \au{Klessen, R.~S}, \au{Iapichino, L.} \& \au{Beattie,
  J.~R}} \yr{2021}  \at{The sonic scale of interstellar turbulence}.
  \jt{Nature Astronomy}  \bvol{5}~(4),  \pg{365--371}.

\bibitem[{Federrath} {\em et~al.\/}(2010){Federrath}, {Roman-Duval}, {Klessen},
  {Schmidt} \& {Mac Low}]{FederrathDuvalKlessenSchmidtMacLow2010}
{\sc \au{{Federrath}, C.}, \au{{Roman-Duval}, J.}, \au{{Klessen}, R.~S.},
  \au{{Schmidt}, W.} \& \au{{Mac Low}, {M.-M.}}} \yr{2010}  \at{{Comparing the
  statistics of interstellar turbulence in simulations and observations.
  Solenoidal versus compressive turbulence forcing}}.  \jt{\aap}  \bvol{512},
  \pg{A81}.

\bibitem[{Federrath} {\em et~al.\/}(2022){Federrath}, {Roman-Duval}, {Klessen},
  {Schmidt} \& {Mac Low}]{FederrathEtAl2022ascl}
{\sc \au{{Federrath}, C.}, \au{{Roman-Duval}, J.}, \au{{Klessen}, R.~S.},
  \au{{Schmidt}, W.} \& \au{{Mac Low}, M.~M.}} \yr{2022} {TG: Turbulence
  Generator}. Astrophysics Source Code Library, record ascl:2204.001,
  \arxiv{arXiv: 2204.001}.

\bibitem[{Fryxell} {\em et~al.\/}(2000){Fryxell}, {Olson}, {Ricker}, {Timmes},
  {Zingale}, {Lamb}, {MacNeice}, {Rosner}, {Truran} \& {Tufo}]{FryxellEtAl2000}
{\sc \au{{Fryxell}, B.}, \au{{Olson}, K.}, \au{{Ricker}, P.}, \au{{Timmes},
  F.~X.}, \au{{Zingale}, M.}, \au{{Lamb}, D.~Q.}, \au{{MacNeice}, P.},
  \au{{Rosner}, R.}, \au{{Truran}, J.~W.} \& \au{{Tufo}, H.}} \yr{2000}
  \at{{FLASH: An Adaptive Mesh Hydrodynamics Code for Modeling Astrophysical
  Thermonuclear Flashes}}.  \jt{\apjs}  \bvol{131},  \pg{273--334}.

\bibitem[{Hatori}(1984)]{Hatori84}
{\sc \au{{Hatori}, T.}} \yr{1984}  \at{{Kolmogorov-style argument for the
  decaying homogeneous MHD turbulence}}.  \jt{J. Phys. Soc. Jpn.}  \bvol{53},
  \pg{2539}.

\bibitem[{Hosking} \& {Schekochihin}(2021)]{HoskingSchekochihin20decay}
{\sc \au{{Hosking}, D.~N.} \& \au{{Schekochihin}, A.~A.}} \yr{2021}
  \at{Reconnection-controlled decay of magnetohydrodynamic turbulence and the
  role of invariants}.  \jt{Phys. Rev. X}  \bvol{11},  \pg{041005}.

\bibitem[{Ishida} {\em et~al.\/}(2006){Ishida}, {Davidson} \&
  {Kaneda}]{Ishida06}
{\sc \au{{Ishida}, T.}, \au{{Davidson}, P.~A.} \& \au{{Kaneda}, Y.}} \yr{2006}
  \at{{On the decay of isotropic turbulence}}.  \jt{J. Fluid Mech.}
  \bvol{564},  \pg{455}.

\bibitem[Kang {\em et~al.\/}(2003)Kang, Chester \& Meneveau]{kang2003decaying}
{\sc \au{Kang, H.~S.}, \au{Chester, S.} \& \au{Meneveau, C.}} \yr{2003}
  \at{Decaying turbulence in an active-grid-generated flow and comparisons with
  large-eddy simulation}.  \jt{Journal of Fluid Mechanics}  \bvol{480},
  \pg{129--160}.

\bibitem[{Kolmogorov}(1941)]{Kolmogorov41c}
{\sc \au{{Kolmogorov}, A.~N.}} \yr{1941}  \at{{Dissipation of energy in locally
  isotropic turbulence}}.  \jt{Dokl. Acad. Nauk SSSR}  \bvol{32},  \pg{16}.

\bibitem[{Loitsyansky}(1939)]{Loitsyansky39}
{\sc \au{{Loitsyansky}, L.~G.}} \yr{1939}  \at{{Some basic laws for isotropic
  turbulent flow}}.  \jt{Trudy Tsentr. Aero.-Gidrodin Inst.}  \bvol{440},
  \pg{3}.

\bibitem[Mac~Low {\em et~al.\/}(1998)Mac~Low, Klessen, Burkert \&
  Smith]{mac1998kinetic}
{\sc \au{Mac~Low, M.-M.}, \au{Klessen, R.~S.}, \au{Burkert, A.} \& \au{Smith,
  M.~D}} \yr{1998}  \at{Kinetic energy decay rates of supersonic and
  super-alfv{\'e}nic turbulence in star-forming clouds}.  \jt{Physical Review
  Letters}  \bvol{80}~(13),  \pg{2754}.

\bibitem[{Matthaeus} \& {Montgomery}(1980)]{MatthaeusMontgomery80}
{\sc \au{{Matthaeus}, W.~H.} \& \au{{Montgomery}, D.}} \yr{1980}
  \at{{Selective decay hypothesis at high mechanical and magnetic Reynolds
  numbers}}.  \jt{Annals of the New York Academy of Sciences}  \bvol{357},
  \pg{203}.

\bibitem[McCullagh(2018)]{McCullagh18}
{\sc \au{McCullagh, P.}} \yr{2018} {\em Tensor Methods in Statistics\/}, 2nd
  edn.  \publ{New York: Dover Publications}.

\bibitem[M{\'e}tais \& Lesieur(1986)]{metais1986statistical}
{\sc \au{M{\'e}tais, O.} \& \au{Lesieur, M.}} \yr{1986}  \at{Statistical
  predictability of decaying turbulence}.  \jt{Journal of the Atmospheric
  Sciences}  \bvol{43}~(9),  \pg{857--870}.

\bibitem[{Moffatt}(1969)]{Moffatt69}
{\sc \au{{Moffatt}, H.~K.}} \yr{1969}  \at{{The degree of knottedness of
  tangled vortex lines}}.  \jt{J. Fluid Mech.}  \bvol{35},  \pg{117}.

\bibitem[{Morillo} \& {Alexakis}(2025)]{MorilloAlexakis25}
{\sc \au{{Morillo}, J.~M.~G.} \& \au{{Alexakis}, A.}} \yr{2025}  \at{{Magnetic
  reconnection, plasmoids and numerical resolution}}.  \jt{J. Fluid Mech.}
  \bvol{1007},  \pg{R3}.

\bibitem[{M{\"u}ller} \& {Biskamp}(2000)]{BiskampMuller00}
{\sc \au{{M{\"u}ller}, W.-C.} \& \au{{Biskamp}, D.}} \yr{2000}  \at{{Scaling
  properties of three-dimensional magnetohydrodynamic turbulence}}.  \jt{Phys.
  Rev. Lett.}  \bvol{84},  \pg{475}.

\bibitem[{Olesen}(1997)]{Olesen97}
{\sc \au{{Olesen}, P.}} \yr{1997}  \at{{Inverse cascades and primordial
  magnetic fields}}.  \jt{Phys. Lett. B}  \bvol{398},  \pg{321}.

\bibitem[Porter {\em et~al.\/}(1994)Porter, Pouquet \&
  Woodward]{porter1994kolmogorov}
{\sc \au{Porter, D.~H}, \au{Pouquet, A} \& \au{Woodward, P.~R}} \yr{1994}
  \at{Kolmogorov-like spectra in decaying three-dimensional supersonic flows}.
  \jt{Physics of Fluids}  \bvol{6}~(6),  \pg{2133--2142}.

\bibitem[{Robertson}(1940)]{robertson1940invariant}
{\sc \au{{Robertson}, H.~P.}} \yr{1940}  \at{The invariant theory of isotropic
  turbulence}.  \jt{Mathematical Proceedings of the Cambridge Philosophical
  Society}  \bvol{36}~(2),  \pg{209}.

\bibitem[{Saffman}(1967)]{saffman67}
{\sc \au{{Saffman}, P.~G.}} \yr{1967}  \at{{The large-scale structure of
  homogeneous turbulence}}.  \jt{J. Fluid Mech.}  \bvol{27},  \pg{581}.

\bibitem[Taylor(1974)]{taylor1974}
{\sc \au{Taylor, J.~B.}} \yr{1974}  \at{Relaxation of toroidal plasma and
  generation of reverse magnetic fields}.  \jt{Phys. Rev. Lett.}  \bvol{33},
  \pg{1139--1141}.

\bibitem[{Taylor} \& {Newton}(2015)]{TaylorNewton15}
{\sc \au{{Taylor}, J.~B.} \& \au{{Newton}, S.~L.}} \yr{2015}  \at{{Special
  topics in plasma confinement}}.  \jt{J. Plasma Phys.}  \bvol{81},
  \pg{205810501}.

\bibitem[{Uzdensky} {\em et~al.\/}(2010){Uzdensky}, {Loureiro} \&
  {Schekochihin}]{Uzdensky10}
{\sc \au{{Uzdensky}, D.~A.}, \au{{Loureiro}, N.~F.} \& \au{{Schekochihin},
  A.~A.}} \yr{2010}  \at{{Fast magnetic reconnection in the plasmoid-dominated
  regime}}.  \jt{Phys. Rev. Lett.}  \bvol{105},  \pg{235002}.

\bibitem[{Vicentin} {\em et~al.\/}(2025){Vicentin}, {Kowal}, {de Gouveia Dal
  Pino} \& {Lazarian}]{Vicentin25}
{\sc \au{{Vicentin}, G.~H.}, \au{{Kowal}, G.}, \au{{de Gouveia Dal Pino},
  E.~M.} \& \au{{Lazarian}, A.}} \yr{2025}  \at{{Do plasmoids induce fast
  magnetic reconnection in well-resolved current sheets in 2D MHD
  simulations?}}  \jt{arXiv:}  \pg{p. arXiv:2510.01060}.

\bibitem[Waagan(2009)]{waagan2009positive}
{\sc \au{Waagan, K.}} \yr{2009}  \at{A positive {MUSCL-Hancock} scheme for
  ideal magnetohydrodynamics}.  \jt{Journal of Computational Physics}
  \bvol{228}~(23),  \pg{8609--8626}.

\bibitem[{Waagan} {\em et~al.\/}(2011){Waagan}, {Federrath} \&
  {Klingenberg}]{WaaganFederrathKlingenberg2011}
{\sc \au{{Waagan}, K.}, \au{{Federrath}, C.} \& \au{{Klingenberg}, C.}}
  \yr{2011}  \at{{A robust numerical scheme for highly compressible
  magnetohydrodynamics: Nonlinear stability, implementation and tests}}.
  \jt{\jcp}  \bvol{230},  \pg{3331--3351}.

\bibitem[{Woltjer}(1958)]{Woltjer58a}
{\sc \au{{Woltjer}, L.}} \yr{1958}  \at{{A theorem on force-free magnetic
  fields}}.  \jt{Proc. Natl. Acad. Sci. U.S.A.}  \bvol{44},  \pg{489}.

\bibitem[{Zhou} {\em et~al.\/}(2022){Zhou}, {Sharma} \&
  {Brandenburg}]{Zhou22_Hosking}
{\sc \au{{Zhou}, H.}, \au{{Sharma}, R.} \& \au{{Brandenburg}, A.}} \yr{2022}
  \at{{Scaling of the Hosking integral in decaying magnetically dominated
  turbulence}}.  \jt{J. Plasma Phys.}  \bvol{88},  \pg{905880602}.

\bibitem[{Zhou} {\em et~al.\/}(2019){Zhou}, {Bhat}, {Loureiro} \&
  {Uzdensky}]{Zhou19}
{\sc \au{{Zhou}, M.}, \au{{Bhat}, P.}, \au{{Loureiro}, N.~F.} \&
  \au{{Uzdensky}, D.~A.}} \yr{2019}  \at{{Magnetic island merger as a mechanism
  for inverse magnetic energy transfer}}.  \jt{Phys. Rev. Res.}  \bvol{1},
  \pg{012004}.

\bibitem[{Zrake}(2014)]{Zrake14}
{\sc \au{{Zrake}, J.}} \yr{2014}  \at{{Inverse cascade of nonhelical magnetic
  turbulence in a relativistic fluid}}.  \jt{Astrophys. J. Lett.}  \bvol{794},
  \pg{L26}.

\end{thebibliography}

\end{document}